\UseRawInputEncoding

\documentclass[prr,twocolumn,superscriptaddress,amsmath,amssymb,longbibliography]{revtex4-2}

\usepackage{graphicx}
\usepackage{latexsym}
\usepackage{amsmath}
\usepackage{amssymb}
\usepackage{amsfonts}
\usepackage{color}
\usepackage{bm}
\usepackage{verbatim}
\usepackage{pagecolor,lipsum}
\usepackage{float}
\usepackage{hyperref}
\usepackage[normalem]{ulem}
\hypersetup{colorlinks=true,urlcolor=blue,linkcolor=blue,citecolor=blue}
\usepackage{comment}
\usepackage{siunitx}

\begin{document}

\title{Mesoscopic superconducting memory based on bistable magnetic textures}

\date{\today}

\author{R. Fermin}
\author{N. Scheinowitz}
\author{J. Aarts}
\affiliation{Huygens-Kamerlingh Onnes Laboratory, Leiden University, P.O. Box 9504, 2300 RA Leiden, The Netherlands.}
\author{K. Lahabi}
\affiliation{Huygens-Kamerlingh Onnes Laboratory, Leiden University, P.O. Box 9504, 2300 RA Leiden, The Netherlands.}\affiliation{Department of Quantum Nanoscience, Kavli Institute of Nanoscience, Delft University of Technology, Delft, 2628 CJ, The Netherlands}

\begin{abstract}
With the ever-increasing energy need to process big data, the realization of low-power computing technologies, such as superconducting logic and memories, has become a pressing issue. Developing fast and non-volatile superconducting memory elements, however, remains a challenge. Superconductor-ferromagnet hybrid devices offer a promising solution, as they combine ultra-fast manipulation of spins with dissipationless readout. Here, we present a new type of non-volatile Josephson junction memory that utilizes the bistable magnetic texture of a single mesoscopic ferromagnet. We use micromagnetic simulations to design an ellipse-shaped planar junction structured from a Nb/Co bilayer. The ellipse can be prepared as uniformly magnetized or as a pair of vortices at zero applied field. The two states yield considerably different critical currents, enabling reliable electrical readout of the element. We describe the mechanism used to control the critical current by applying numerical calculations to quantify the local stray field from the ferromagnet, which shifts the superconducting interference pattern. By combining micromagnetic modeling with bistable spin-textured junctions, our approach presents a novel route towards realizing superconducting memory applications.
\end{abstract}

 \maketitle
 
\section{Introduction}

In recent years, energy-hungry data centers have accounted for more than 1\% of the global electricity consumption; this number is projected to increase to 3-13\% by 2030.\cite{Andrae2015} As data centers are already responsible for 0.3\% of the overall carbon emissions, the development of low-power supercomputers has become an immediate global concern.\cite{Jones2018} Due to their non-dissipative nature, superconducting logic and memory devices present a tantalizing route to address this issue, promising enormous savings, even accounting for cryogenic cooling.\cite{Holmes2013,Soloviev2017,Chen2020,Butters2021,Hilgenkamp2020} While several classes of high-performance superconducting processors already exist, cryogenic memories remain relatively underdeveloped and far from meeting their applications. Non-volatility and scalability, in particular, have been the major obstacles in the industrial realization of superconducting memories.
 \begin{figure}[htb!]
 \centerline{$
 \begin{array}{c}
  \includegraphics[width=0.9\linewidth]{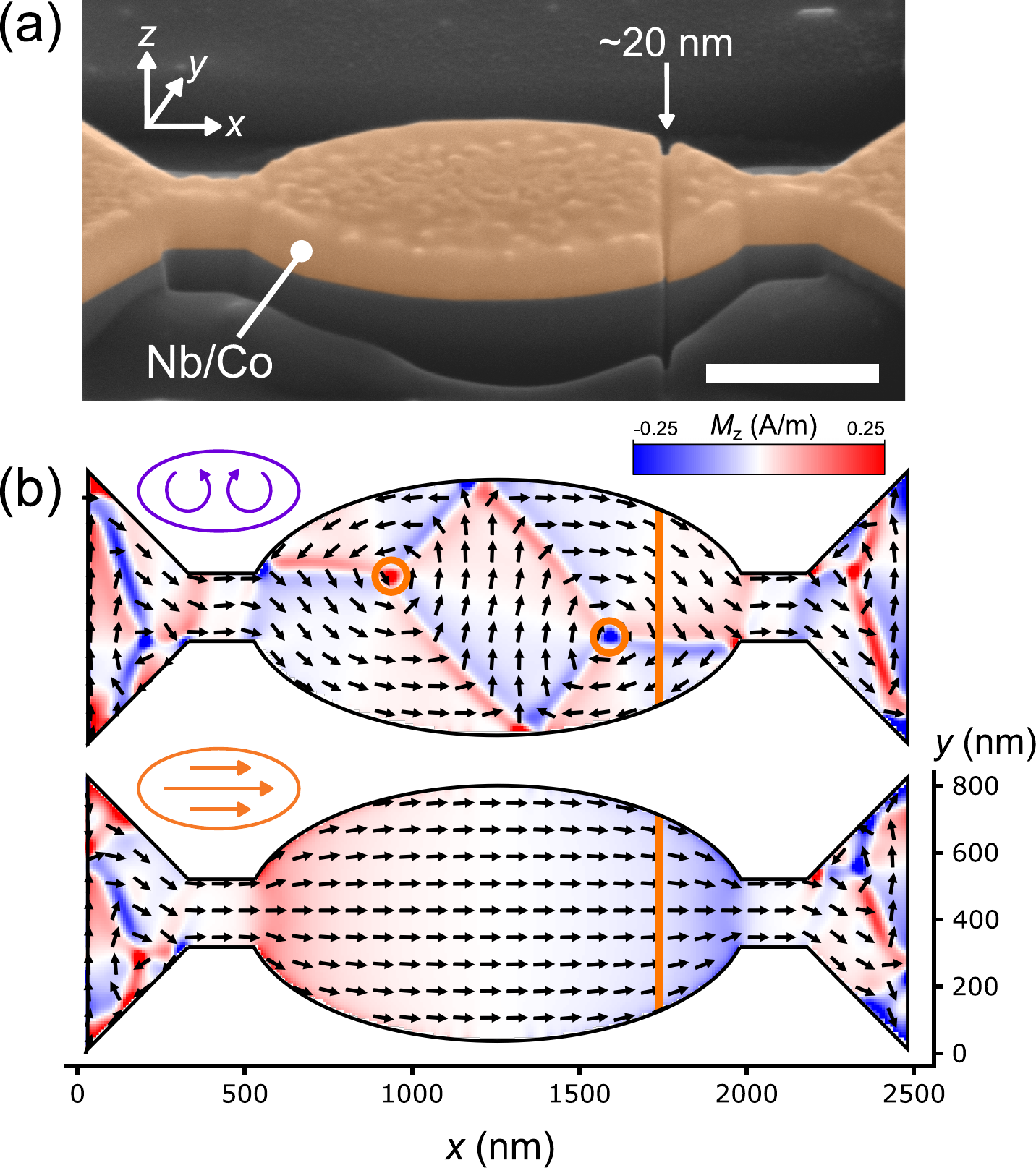}
 \end{array}$}
 \caption{(a) False colored scanning electron micrograph of the ellipse-shaped SFS junction device A. The Co weak link is formed by the indicated trench ($\backsim$20 nm in size) that separates the two Nb electrodes. The scale bar corresponds to 500~nm. (b) Shows the simulated bistable spin textures at zero field. The cobalt is either magnetized along the long axis of the device (M-state), or hosts two stable vortices (V-state). The vortex cores are highlighted by orange circles and the trench location is indicated by the orange line. The color scale indicates the out-of-plane magnetization of the top of the cobalt layer.}\label{fig1}
 \end{figure}

The combination of superconducting (S) and ferromagnetic (F) elements can tackle the non-volatility issue through SFS devices, in which different (stable) magnetic configurations yield different critical supercurrents. Much work has been done on stacked magnetic Josephson junctions with a weak ferromagnetic barrier, where small magnetic field pulses could switch the critical current.~\cite{Larkin2012,Vernik2013,Caruso2018,Caruso2018a} However, the weak ferromagnet used in such systems has a typically low Curie temperature (around 10~K), limiting their non-volatility. Josephson memories have also been extensively studied in so-called pseudo-spin-valve (SFNF$'$S) devices, where the magnetization of two ferromagnetic layers, separated by a normal metal, could be set parallel or antiparallel to yield a change in critical current.~\cite{Baek2014, Golod2015, Dayton2018, Nevirkovets2018, Niedzielski2018, Madden2018,  Satchell2020} The work on pseudo-spin-valves has also been extended to multilayer Josephson devices featuring triplet supercurrents, which not only carry charge but also spin.~\cite{Martinez2016, Glick2018} Finally, all-oxide devices, which can operate at high temperatures were recently examined.~\cite{DeAndresPrada2019} In general, the existing non-volatile Josephson memories utilize an intricate layer set and face a number of obstacles, such as interlayer coupling between magnetic layers and stochastic distribution of magnetic flux in the junction. Other problems include, insufficient contrast between states for electrical readout, and the requirement for applying a large writing field.

To address these challenges, we present an alternative route for developing non-volatile Josephson memory elements, where information is stored by the spin texture of a single mesoscopic ferromagnet. Magnetic textures in general, and specifically ferromagnetic vortices, can be manipulated by the application of pulsed microwave radiation, leading to the realization of ultra-fast electronics.\cite{Weigand2009,Uhlir2013,Kammerer2011} Combining information inscription in spin textures with the non-dissipative nature of supercurrents, can thus lead to the realization of high-speed, low-power, and non-volatile memory elements. The synergy between superconductivity and magnetic textures is amplified further by allowing for triplet supercurrents, paving the road for superconducting spintronics.\cite{Linder2015, Eschrig2008, Robinson2021} Textured SF-hybrid devices, therefore, open a truly new paradigm in superconducting memory applications.

Here we show the realization of a micrometer-sized superconducting memory element based on bistable spin textures in the F-layer of an elliptical planar SFS Josephson junction. By combining micromagnetic simulations with nanostructured hybrid devices, we are able to control the transport behaviour of our devices with their magnetic texture. By applying a relatively small magnetic field of 40 mT for the writing operation, we can reliably switch between two stable spin textures: one is uniformly magnetized, and the other hosts a pair of ferromagnetic vortices. These correspond to minimum and maximum critical supercurrent of the junction, respectively. We obtain a factor of five difference in the critical current between the two states, which allows for a facile and reliable electrical readout of the element. By quantifying the local stray field from the ferromagnet, we show that the difference in critical current is caused by a shift in the superconducting interference pattern. Moreover, the magnetic bistability enables us to use relatively small writing fields to control a considerably larger local stray field from the ferromagnet. We confirm that the memory is stable, non-volatile, and non-destructive upon readout.

 \begin{figure}[t!]
 \centerline{$
 \begin{array}{c}
  \includegraphics[width=0.9\linewidth]{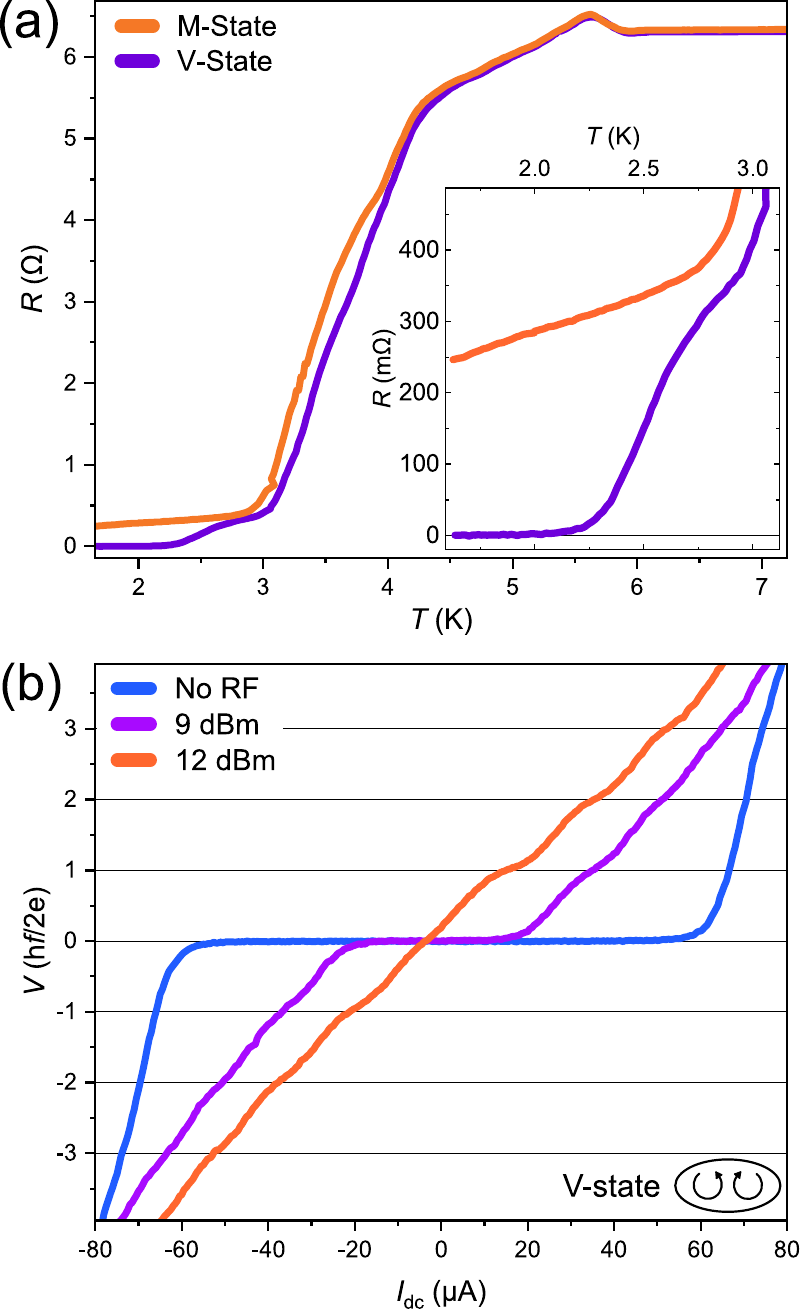}
 \end{array}$}
 \caption{Basic transport characteristics of device A. In (a) we show the resistance versus temperature curves for both the M- and V-state, measured using 20~$\mu$A current bias. The critical current is suppressed below 20~$\mu$A in the M-state. The inset shows the low-temperature behavior. (b) depicts three $IV$-characteristics of the sample in the V-state, acquired under different RF radiation powers. The frequency is 1.6 GHz and the temperature is 1.6 K. The measured voltage is normalized to $hf /2e$, and shows a clear Shapiro response.}\label{fig2} 
 \end{figure}

\section{Results and discussion}

\subsection{Ellipse-shaped SFS junctions}

 \begin{figure*}[t!]
 \centerline{$
 \begin{array}{c}
  \includegraphics[width=0.95\linewidth]{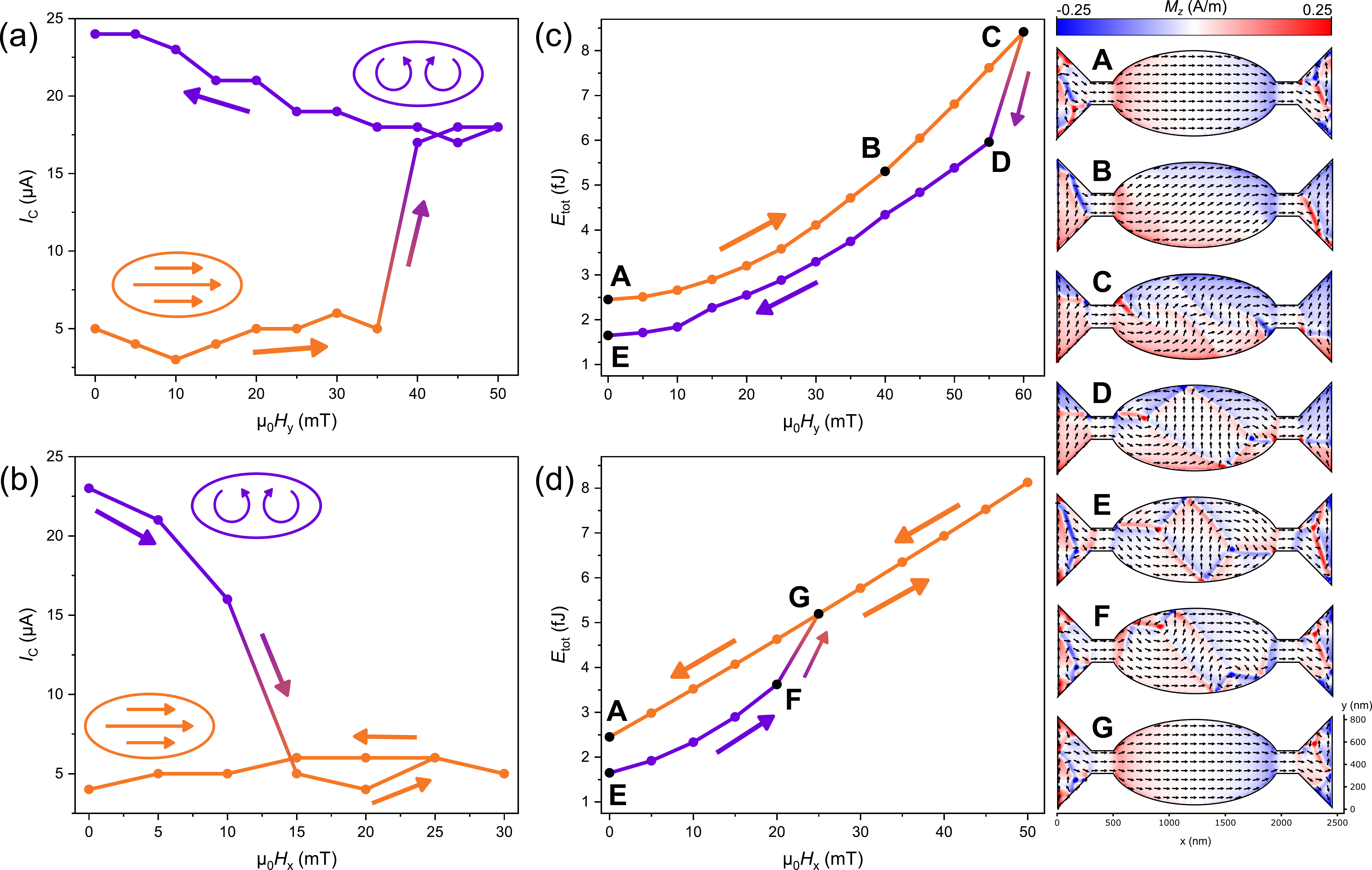}
 \end{array}$}
 \caption{The critical current $I_{\text{c}}$, the total simulated energy $E_{tot}$ and the spin texture of the device as a function of H$_x$ (field along the long axis), and $H_y$ (along the short axis). Results obtained on device B at 1.7 K. (a) and (b) depict $I_{\text{c}}(H_y)$ and $I_{\text{c}}(H_x)$. In (a) the field is increased along the short axis of the ellipse. At 40~mT $I_\text{c}$ increases sharply, effectively transforming the M-state to the V-state. The reverse operation is shown in (b): by increasing the field - parallel to the long axis of the ellipse - to 15 mT, the critical current drops to 5~$\mu$A, bringing the device back to the M-state. (c) and (d) show $E_{tot}$ corresponding to the in-plane field sweeps. In both panels, two branches can be seen, corresponding to the V- and M-state. A change of magnetic state coincides with a jump from one branch to the other. We are able to correctly predict the switching fields to within a small error. The right-hand column contains snapshots of the spin texture during the simulated field sweeps. Here the color scale indicates the magnetization of the top of the Co-layer.}\label{fig3}
 \end{figure*}   

A cobalt (65 nm) niobium (50 nm) bilayer is deposited in an ultra-high vacuum by Ar-sputtering on a four-probe geometry lift-off pattern. Elliptic SFS devices are structured from the Co/Nb bilayer using focused ion beam milling. Figure \ref{fig1} shows a false-colored scanning electron micrograph of such an ellipse device, having dimensions of 1500~nm by 750~nm (long and short axes, respectively). Micromagnetic simulations show that the zero-field spin texture of such devices is expected to be bistable: either the cobalt layer is magnetized along the long axis of the ellipse (we call this the M-state) or two ferromagnetic vortices are stable near each focal point of the ellipse (V-state). We depict simulation results from these states in Figure \ref{fig1}b. The full technical details of the simulations are presented in Appendix A. Essentially, this allows for writing a bit with the value "0" (M-state) or "1" (V-state) into the spin texture of the device. We prepare the states by first applying a 40 mT in-plane field either parallel to the long or short axis of the devices. Next, after removing the field along the short (long) axis, the V-state (M-state) is stable in the device. 

Using ultra-low ion-beam currents, We fabricate a planar Josephson junction in the ellipse by locally removing the Nb top-layer. This creates an approximately ~20 nm wide trench, which effectively separates the Nb contacts and forces the supercurrent through the ferromagnet. Planar SFS-junctions fabricated in this manner were used to study triplet supercurrents in previous works.\cite{Lahabi2017a,Co_disk_paper} The trench is positioned near the focal point of the ellipse, which corresponds to the approximate location of one of the vortex cores in the V-state. As discussed later, this particular location is designed to optimize the stray field-driven mechanism used to assign different critical currents to the M- and V-states. In this paper, we describe two of these elliptical devices (devices A and B), although all observations are repeated for a large number of samples. The sample resistance is measured using a DC nanocurrent source (Keithley 6221) and a nanovolt meter (Keithley 2182A) in a 4-probe fashion, where the voltage drop over both the bar-shaped contacts and the ellipse is measured. Figure \ref{fig2} shows the resistance ($R$) versus temperature ($T$) of device A, measured with a current of 20~$\mu$A. If the sample is prepared in the V-state, we find two transitions. The first corresponds to the bulk superconducting transition temperature ($T_{\text{c}}$) of the Nb layer, the second is due to the superconducting proximity effect in the SFS junction formed by the trench. In contrast, the M-state does not show a second transition, indicating that the critical current ($I_{\text{c}}$) of the junction in this state is suppressed to below the measurement current. This difference in $I_{\text{c}}$ between the two states allows us to electrically readout the magnetic state of the memory element. To unambiguously demonstrate the superconducting proximity effect in our devices, we show the appearance of Shapiro steps in the current-voltage ($IV$-)characteristic of device A under the application of RF radiation, which is supplied by a nearby antenna. We carry out these measurements at 1.6 K in the V-state.

\clearpage

\begin{figure}[htb!]
\centerline{$
\begin{array}{c}
 \includegraphics[width=0.9\linewidth]{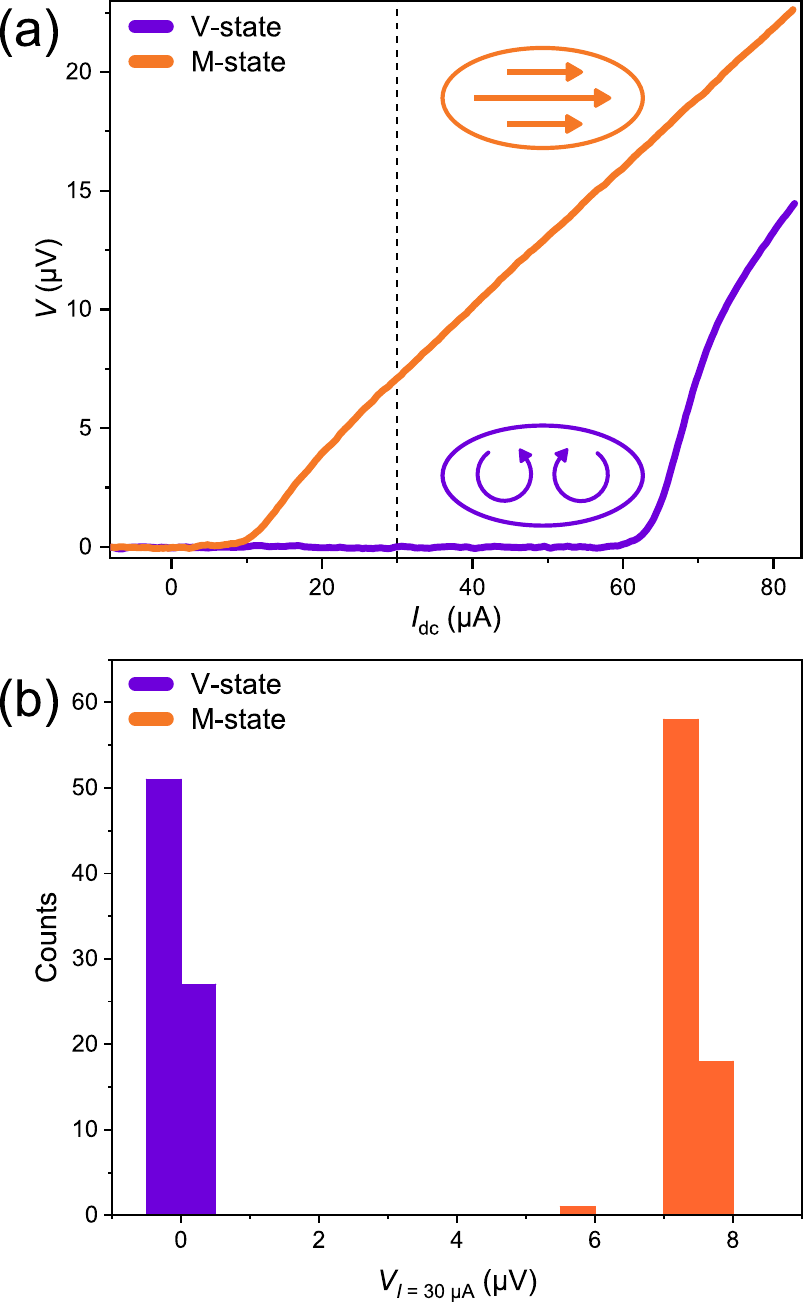}
\end{array}$}
\caption{Fidelity of the writing cycles of device A. The $IV$-characteristics of the two states, measured at 1.6 K, are shown in (a). Based on these curves, we define a readout current that corresponds to a finite voltage in the M-state and zero voltage in the V-state. We cycle the device between the two states 78 times and measure the voltage before each subsequent switch. (b) shows a histogram of the readout voltages, measured using a 30~$\mu$A current (the dotted line in (a)).}\label{fig4}
\end{figure}

\noindent The step height in the $IV$-characteristic shown in Figure \ref{fig2}b is in units $hf/2e$ (where $h$ is the Planck constant, $f$ is the frequency of the RF radiation, and $e$ the electron charge), confirming the Shapiro response.

\subsection{Controllable switching}

To show the bistability of the spin textures, we examine the dependence of $I_{\text{c}}$ on the in-plane 'writing' fields H$_x$ and H$_y$ along the long and the short axis of the ellipse, respectively. The results are shown in Figure \ref{fig3}a,b, accompanied by the simulated spin textures. If the sample is in the M-state ("0"), $I_{\text{c}}$ is 5~$\mu$A. The datum "1" can be written by increasing the field along the short axis of the ellipse: at a field of 40~mT, $I_{\text{c}}$ grows to above 20~$\mu$A, which is accompanied by a buckling of the simulated spin texture. Upon decreasing the field to zero, the high $I_{\text{c}}$ state is retained and the buckled spin texture evolves into two vortices. Therefore, the device is converted into the V-state ("1"). Conversely, the M-state can be written in the device by applying a field parallel to the long axis of the ellipse. The field effectively displaces the vortices towards the edges of the sample. At a field of 15~mT, they are fully pushed out from the ellipse and the spins align in the M-state. This is accompanied by a drop in $I_{\text{c}}$ back to 5~$\mu$A. The low $I_{\text{c}}$ is stable upon decreasing the field back to zero in the M-state.

Figure~\ref{fig3}c,d show the total energy associated with the simulated spin texture as a function of externally applied field (i.e., the sum of the exchange, demagnetization, and Zeeman energies). Two branches can be discerned corresponding to the two states, both stable at zero field. In the simulations, the buckled state, which is required for the stabilization of the V-state state, appears above 45 mT. During the converse switching operation, a field of 25 mT is required to magnetize the ellipse. Remarkably, we find both switching fields closely follow the experimental values with only a small offset, which can be attributed to the reduced geometry of the contacts, and the finite field steps used in the simulations.

We examine the performance of our devices by repeatably switching them between the two states while studying the transport behavior. Figure~\ref{fig4}a shows the zero-field $IV$-characteristic of the two states. Based on the difference in $I_{\text{c}}$, we can define a readout current that corresponds to a finite voltage in the M-state, but a zero voltage in the V-state. We cycle the device between the two states (using a 40~mT field) and acquire the voltage at the readout current in each state. In Figure \ref{fig4}b we show a histogram of the measured voltage at a readout current of 30~$\mu$A, totaling 78 subsequent writing cycles. There is no overlap in the histogram, making the states highly distinguishable. As the readout current density is far too low to alter the spin texture, the readout operation is fully non-destructive. Furthermore, by heating the ellipse to above the $T_{\text{c}}$ of the superconducting layer between subsequent write and read operations, we confirm the non-volatility of its memory. This was found to hold, even when the sample is warmed to room temperature and stored for days. Additionally, to test the resilience to perturbations of the spin texture, we try to cycle between the states using a field of 10~mT. We find these fields insufficient to alter $I_{\text{c}}$, demonstrating the stability of our devices.

 \begin{figure*}[tb]
 \centerline{$
 \begin{array}{c}
  \includegraphics[width=0.95\linewidth]{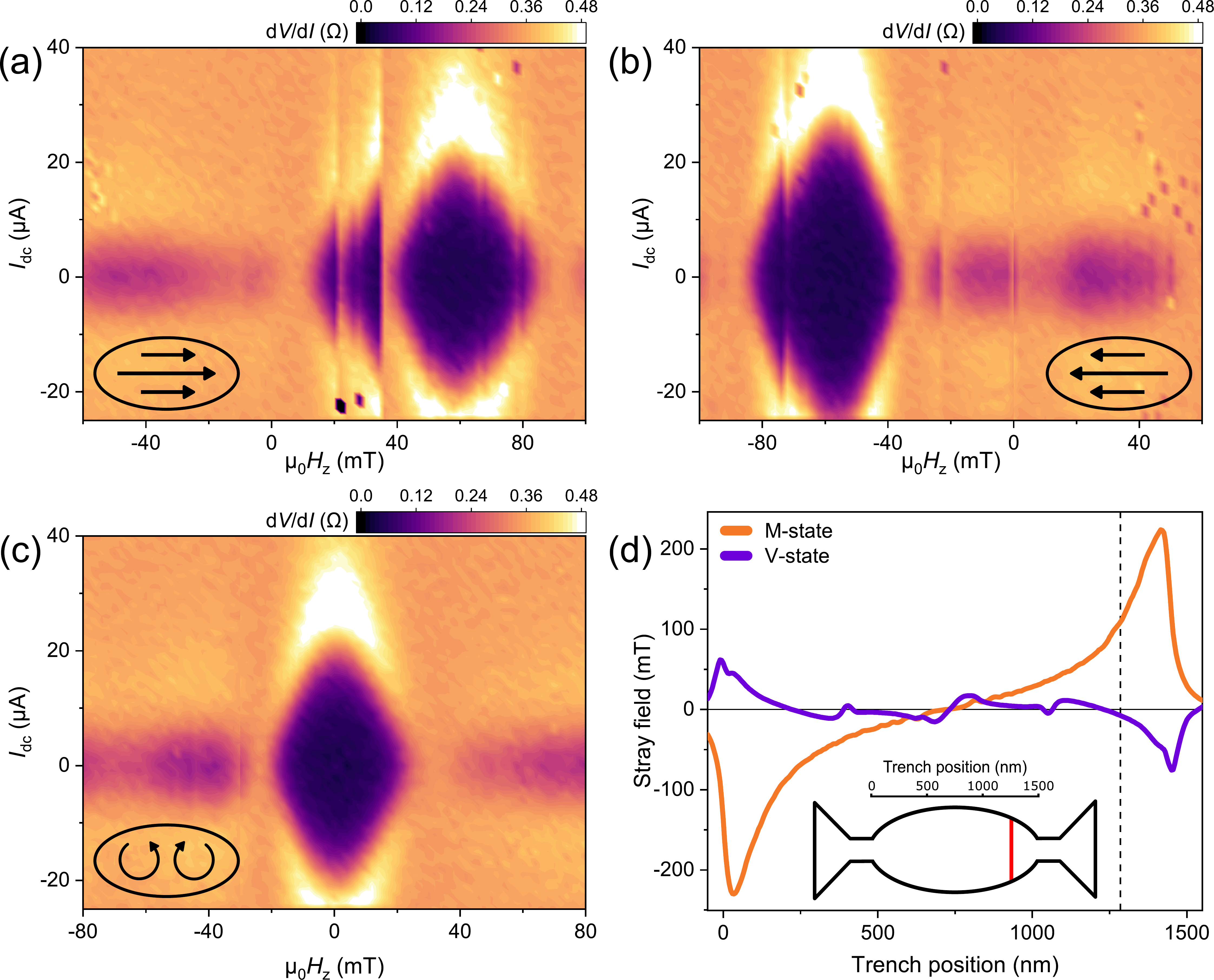}
 \end{array}$}
 \caption{Stray fields cause a shift in I$_{\text{c}}$. (a) to (c) respectively show the superconducting interference (SQI) patterns in the right magnetized M-state, left magnetized M-state, and the V-state obtained on device B. Although the shape of the SQI pattern does not change, the center lobe of the pattern is shifted in the V-state with respect to the pattern in the M-state (for both magnetizations). Some measurement glitches are visible in (a) and (b). These result from sudden small magnetic configurations in the sample. The shift is caused by stray fields from the ferromagnetic layer, as is demonstrated in (d): the simulated field shift for different trench locations. The vertical line in (d) corresponds to the actual location of the trench (see Figure \ref{fig1}). The inset shows the geometry of the sample with the corresponding scale distribution. The red line indicates the actual location of the trench.}\label{fig5}
 \end{figure*}   

\subsection{Stray field driven mechanism}

We now discuss the mechanism used for suppressing the $I_{\text{c}}$ in the M-state, where we utilize the local stray field from the ferromagnet itself. To illustrate this, we perform superconducting quantum interference (SQI) measurements, where we sweep the out-of-plane field and acquire an $IV$-characteristic at each field. During the SQI measurements, we obtain a direct measure of $\text{d}V/\text{d}I$, by obtaining an AC-voltage using a Synktek MCL1-540 multichannel lock-in. For a regular junction, the interference pattern $I_{\text{c}}(B)$ corresponds to the well-known Fraunhofer pattern. Note that the magnitude of the out-of-plane fields is insufficient to alter the global spin texture of the device (i.e., the out-of-plane fields cannot transform the M-state to the V-state or vice versa). We present such SQI measurements, obtained on sample B, for the V-state, as well as for the left and right magnetized M-state in the color maps of Figure \ref{fig5}a-c. For this sample, we observe a pattern with a single middle lobe and heavily suppressed side lobes.~\footnote{Samples with higher $I_{\text{c}}$ values do show a Fraunhofer interference pattern, indicating a uniform distribution of $I_{\text{c}}$ throughout the weak link.} For the V-state (Figure~\ref{fig5}c), the pattern is centered at zero field. By switching to either of the M-states (Figure \ref{fig5}a,b), the shape of the SQI pattern is not altered, yet clearly shifted by around 60 mT from zero applied field. Moreover, the magnetization direction in the M-state determines the sign of this shift. Therefore, it is evident that the suppressed $I_{\text{c}}$ of the M-state is a result of the local stray fields from the Co layer, introducing a shift in the $I_{\text{c}}(B)$ pattern. However, what makes our spin-textured devices truly stand out from previous Josephson memory elements, is that we use a writing field (40 mT) that is noticeably smaller than the resulting shift (60 mT). 

In the following, we describe how the pattern shift  can be quantified by simulating the stray flux entering the junction. We sum the z-component of the flux - evaluated in a thin sheet above the Cobalt layer - over an integration window to obtain the local stray fields threading area of the window. The size of this window is chosen to reflect the trench area, therefore yielding an estimate of the stray fields penetrating the junction. Figure \ref{fig5}d displays the simulated (stray field-induced) pattern shift as a function of the window location; the vertical line indicates the location of the trench in the actual sample. This show that the V-state corresponds to negligible stray fields at the junction, accounting for the centered pattern in Figure \ref{fig5}c with no offset. The M-state corresponds to an 80 mT shift, which closely follows the 60 mT offset in the SQI patterns. Note that the simulated stray field is symmetric with respect to the center of the sample. This reflects the sign change that occurs when the magnetization direction is reversed.

Note that the spin texture snapshots of Figure \ref{fig3} show the source of the stray fields. In the M-state, the ellipse features a dipole-like field along its long axis, which is absent in the V-state. We do observe some stray fields in the V-state, however these average out over the junction area. Therefore the total shift in the SQI pattern remains zero. During the field sweep along the short axis of the ellipse (i.e., switch from M-state to V-state), the spin texture buckles before two vortices stabilize, as can be seen in the snapshots B and C in Figure \ref{fig3}. When this buckling occurs, the dipole-like field no longer points along the long axis the ellipse and consequently the stray fields average out over the junction area. This explains why the high $I_{\text{c}}$ state occurs before two vortices enter in the ellipse.

\subsection{Pairing symmetry of the supercurrents}

In previous work on similar SFS junctions with a circular geometry - containing a single ferromagnetic vortex - we found spin texture to be a prerequisite for the superconducting proximity effect, which led to the conclusion that the transport in these disk-shaped SFS junctions, is dominated by long-range triplet (LRT) Cooper pairs.\cite{Lahabi2017a,Co_disk_paper} Contrary to conventional Cooper pairs, these LRT pairs consist of spin-polarized electrons that can penetrate a ferromagnet over far longer length scales (up to hundreds of nm in a half-metal).\cite{Singh2016,Keizer2006,Anwar2010,Eschrig2008}

In this case however, we cannot be certain about the LRT scenario. Due to the similarity to the circular devices, LRT correlations can be expected in the V-state of the ellipse-shaped devices described here. However, there are a number of inconsistencies, the most notable ones are discussed here. Firstly, contrary to the disk-shaped devices, removing the spin texture from our junctions does not suppress the proximity effect, i.e., the maximum of $I_{\text{c}}(B)$ for both the M- and V-states are quite similar. Secondly, while the SQI patterns of the disk-shaped devices are highly sensitive to the subtle changes in the spin texture, the elliptic junctions show the same type of pattern for widely different magnetic states. Both of these observations seem to counter the notion of the LRT correlations, and indicate that transport in the elliptic junctions is carried by the short-range triplet correlations instead (i.e., Cooper pairs with no spin projection). However, one could argue that even the M-state still features some spin texture near the edges of the sample (due to the tapered shape of the ellipse; see Figure \ref{fig1}b), which possibly allows for the generation of LRT correlations. Finally, the similar SQI patterns between the magnetic states can possibly be caused by the relatively short junction width.\footnote{The length of the trench in the disk-shaped devices is typically between 1~$\mu$m and 1.6~$\mu$m, whereas the trench length of the ellipse devices presenter here is less than 600 nm.} At this stage, no conclusions can be drawn on the triplet or singlet nature of the supercurrents in our devices, and further experiments are necessary. We emphasize, however, that the mechanism for the shifting in the SQI patterns (i.e., the stray fields) hold for both singlet and triplet pairing symmetries.

\section{Conclusion and outlook}

We demonstrate the realization of a non-volatile superconducting memory element based on magnetic textures in an SFS Josephson junction. The ellipsoidal shape of the device leads to two stable magnetic states at zero applied magnetic field: a fully magnetized state and a state containing ferromagnetic vortices. These can be associated with the data "0" and "1" respectively. By applying and removing an externally applied in-plane field, we can reliably switch between these states, performing a write operation. We find a strong suppression of the critical current in the magnetized state, which allows for a non-destructive determination of the spin state, and therefore electrical readout of the bit. By combining transport experiments and micromagnetic simulations, we show that the difference in critical current between the two states results from internal stray fields originating from the ferromagnetic layer of the device, which creates a local offset-field in the junction. Since the ferromagnetic textures can be controllably manipulated with RF radiation, storing information in mesoscopic SF devices can provide an exciting avenue for realizing ultra-fast electronics. Besides, strain-mediated switching of similar elliptical devices has been recently theoretically studied, and shown to be a good alternative to switching using externally applied fields.\cite{Song2022} Combined with the possibility of long-range triplet supercurrents in hybrid devices, the memory devices presented here can form an essential building block for superconducting spintronics.

\section{Acknowledgements}
This work was supported by the Dutch Research Council (NWO) as part of the Frontiers of Nanoscience (NanoFront) program, and through NWO Projectruimte grant 680.91.128. The work was also supported by EU Cost Action CA16218 (NANOCOHYBRI) and benefited from access to the Netherlands Centre for Electron Nanoscopy (NeCEN) at Leiden University.

\vspace{4mm}

\appendix

\section{Micromagnetic modeling}

Micromagnetic simulations are performed using the GPU-based MuMax$^3$ program.\cite{Vansteenkiste2014} The cobalt layer is divided into a mesh of 5 nm cubic cells. The exchange coefficient is set to 3 $\times~10^{-11}$ J/m and the saturation magnetization is 1.4  $\times~10^{-6}$ A/m. We choose the Gilbert damping constant to be 0.5, to allow for faster convergence and therefore increased computation speed. As the cobalt film is polycrystalline due to the Ar-sputtering method, we use a zero anisotropy constant. To evaluate the local stray fields emerging from the ferromagnet, we simulate vacuum cells above the cobalt layer. These cells feature zero spin, however, they are threaded by a magnetic field flux. Summing the z-component of the flux in the vacuum layer over an integration window leads to an expression of the local stray field. The area of the integration window is determined by the local width of the ellipse, and a fixed length of 20~nm. This reflects the size of the trench.


\begin{thebibliography}{36}%
\makeatletter
\providecommand \@ifxundefined [1]{%
 \@ifx{#1\undefined}
}%
\providecommand \@ifnum [1]{%
 \ifnum #1\expandafter \@firstoftwo
 \else \expandafter \@secondoftwo
 \fi
}%
\providecommand \@ifx [1]{%
 \ifx #1\expandafter \@firstoftwo
 \else \expandafter \@secondoftwo
 \fi
}%
\providecommand \natexlab [1]{#1}%
\providecommand \enquote  [1]{``#1''}%
\providecommand \bibnamefont  [1]{#1}%
\providecommand \bibfnamefont [1]{#1}%
\providecommand \citenamefont [1]{#1}%
\providecommand \href@noop [0]{\@secondoftwo}%
\providecommand \href [0]{\begingroup \@sanitize@url \@href}%
\providecommand \@href[1]{\@@startlink{#1}\@@href}%
\providecommand \@@href[1]{\endgroup#1\@@endlink}%
\providecommand \@sanitize@url [0]{\catcode `\\12\catcode `\$12\catcode
  `\&12\catcode `\#12\catcode `\^12\catcode `\_12\catcode `\%12\relax}%
\providecommand \@@startlink[1]{}%
\providecommand \@@endlink[0]{}%
\providecommand \url  [0]{\begingroup\@sanitize@url \@url }%
\providecommand \@url [1]{\endgroup\@href {#1}{\urlprefix }}%
\providecommand \urlprefix  [0]{URL }%
\providecommand \Eprint [0]{\href }%
\providecommand \doibase [0]{https://doi.org/}%
\providecommand \selectlanguage [0]{\@gobble}%
\providecommand \bibinfo  [0]{\@secondoftwo}%
\providecommand \bibfield  [0]{\@secondoftwo}%
\providecommand \translation [1]{[#1]}%
\providecommand \BibitemOpen [0]{}%
\providecommand \bibitemStop [0]{}%
\providecommand \bibitemNoStop [0]{.\EOS\space}%
\providecommand \EOS [0]{\spacefactor3000\relax}%
\providecommand \BibitemShut  [1]{\csname bibitem#1\endcsname}%
\let\auto@bib@innerbib\@empty
\bibitem [{\citenamefont {Andrae}\ and\ \citenamefont
  {Edler}(2015)}]{Andrae2015}%
  \BibitemOpen
  \bibfield  {author} {\bibinfo {author} {\bibfnamefont {A.}~\bibnamefont
  {Andrae}}\ and\ \bibinfo {author} {\bibfnamefont {T.}~\bibnamefont {Edler}},\
  }\bibfield  {title} {\bibinfo {title} {{On Global Electricity Usage of
  Communication Technology: Trends to 2030}},\ }\href
  {https://doi.org/10.3390/challe6010117} {\bibfield  {journal} {\bibinfo
  {journal} {Challenges}\ }\textbf {\bibinfo {volume} {6}},\ \bibinfo {pages}
  {117} (\bibinfo {year} {2015})}\BibitemShut {NoStop}%
\bibitem [{\citenamefont {Jones}(2018)}]{Jones2018}%
  \BibitemOpen
  \bibfield  {author} {\bibinfo {author} {\bibfnamefont {N.}~\bibnamefont
  {Jones}},\ }\bibfield  {title} {\bibinfo {title} {{How to stop data centres
  from gobbling up the world's electricity}},\ }\href
  {https://doi.org/10.1038/d41586-018-06610-y} {\bibfield  {journal} {\bibinfo
  {journal} {Nature}\ }\textbf {\bibinfo {volume} {561}},\ \bibinfo {pages}
  {163} (\bibinfo {year} {2018})}\BibitemShut {NoStop}%
\bibitem [{\citenamefont {Holmes}\ \emph {et~al.}(2013)\citenamefont {Holmes},
  \citenamefont {Ripple},\ and\ \citenamefont {Manheimer}}]{Holmes2013}%
  \BibitemOpen
  \bibfield  {author} {\bibinfo {author} {\bibfnamefont {D.~S.}\ \bibnamefont
  {Holmes}}, \bibinfo {author} {\bibfnamefont {A.~L.}\ \bibnamefont {Ripple}},\
  and\ \bibinfo {author} {\bibfnamefont {M.~A.}\ \bibnamefont {Manheimer}},\
  }\bibfield  {title} {\bibinfo {title} {{Energy-Efficient Superconducting
  Computing—Power Budgets and Requirements}},\ }\href
  {https://doi.org/10.1109/tasc.2013.2244634} {\bibfield  {journal} {\bibinfo
  {journal} {IEEE Trans. Appl. Supercond.}\ }\textbf {\bibinfo {volume} {23}},\
  \bibinfo {pages} {1701610} (\bibinfo {year} {2013})}\BibitemShut {NoStop}%
\bibitem [{\citenamefont {Soloviev}\ \emph {et~al.}(2017)\citenamefont
  {Soloviev}, \citenamefont {Klenov}, \citenamefont {Bakurskiy}, \citenamefont
  {Kupriyanov}, \citenamefont {Gudkov},\ and\ \citenamefont
  {Sidorenko}}]{Soloviev2017}%
  \BibitemOpen
  \bibfield  {author} {\bibinfo {author} {\bibfnamefont {I.~I.}\ \bibnamefont
  {Soloviev}}, \bibinfo {author} {\bibfnamefont {N.~V.}\ \bibnamefont
  {Klenov}}, \bibinfo {author} {\bibfnamefont {S.~V.}\ \bibnamefont
  {Bakurskiy}}, \bibinfo {author} {\bibfnamefont {M.~Y.}\ \bibnamefont
  {Kupriyanov}}, \bibinfo {author} {\bibfnamefont {A.~L.}\ \bibnamefont
  {Gudkov}},\ and\ \bibinfo {author} {\bibfnamefont {A.~S.}\ \bibnamefont
  {Sidorenko}},\ }\bibfield  {title} {\bibinfo {title} {{Beyond Moore's
  technologies: Operation principles of a superconductor alternative}},\ }\href
  {https://doi.org/10.3762/bjnano.8.269} {\bibfield  {journal} {\bibinfo
  {journal} {Beilstein J. Nanotechnol.}\ }\textbf {\bibinfo {volume} {8}},\
  \bibinfo {pages} {2689} (\bibinfo {year} {2017})}\BibitemShut {NoStop}%
\bibitem [{\citenamefont {Chen}\ \emph {et~al.}(2020)\citenamefont {Chen},
  \citenamefont {Wu}, \citenamefont {Wang}, \citenamefont {Pan}, \citenamefont
  {Zhang}, \citenamefont {Zeng}, \citenamefont {Liu}, \citenamefont {Ma},
  \citenamefont {Peng}, \citenamefont {Wang}, \citenamefont {Ren},\ and\
  \citenamefont {Wang}}]{Chen2020}%
  \BibitemOpen
  \bibfield  {author} {\bibinfo {author} {\bibfnamefont {L.}~\bibnamefont
  {Chen}}, \bibinfo {author} {\bibfnamefont {L.}~\bibnamefont {Wu}}, \bibinfo
  {author} {\bibfnamefont {Y.}~\bibnamefont {Wang}}, \bibinfo {author}
  {\bibfnamefont {Y.}~\bibnamefont {Pan}}, \bibinfo {author} {\bibfnamefont
  {D.}~\bibnamefont {Zhang}}, \bibinfo {author} {\bibfnamefont
  {J.}~\bibnamefont {Zeng}}, \bibinfo {author} {\bibfnamefont {X.}~\bibnamefont
  {Liu}}, \bibinfo {author} {\bibfnamefont {L.}~\bibnamefont {Ma}}, \bibinfo
  {author} {\bibfnamefont {W.}~\bibnamefont {Peng}}, \bibinfo {author}
  {\bibfnamefont {Y.}~\bibnamefont {Wang}}, \bibinfo {author} {\bibfnamefont
  {J.}~\bibnamefont {Ren}},\ and\ \bibinfo {author} {\bibfnamefont
  {Z.}~\bibnamefont {Wang}},\ }\bibfield  {title} {\bibinfo {title}
  {{Miniaturization of the superconducting memory cell via a three-dimensional
  Nb nano-superconducting quantum interference device}},\ }\href
  {https://doi.org/10.1021/acsnano.0c04405} {\bibfield  {journal} {\bibinfo
  {journal} {ACS Nano}\ }\textbf {\bibinfo {volume} {14}},\ \bibinfo {pages}
  {11002} (\bibinfo {year} {2020})}\BibitemShut {NoStop}%
\bibitem [{\citenamefont {Butters}\ \emph {et~al.}(2021)\citenamefont
  {Butters}, \citenamefont {Baghdadi}, \citenamefont {Onen}, \citenamefont
  {Toomey}, \citenamefont {Medeiros},\ and\ \citenamefont
  {Berggren}}]{Butters2021}%
  \BibitemOpen
  \bibfield  {author} {\bibinfo {author} {\bibfnamefont {B.~A.}\ \bibnamefont
  {Butters}}, \bibinfo {author} {\bibfnamefont {R.}~\bibnamefont {Baghdadi}},
  \bibinfo {author} {\bibfnamefont {M.}~\bibnamefont {Onen}}, \bibinfo {author}
  {\bibfnamefont {E.~A.}\ \bibnamefont {Toomey}}, \bibinfo {author}
  {\bibfnamefont {O.}~\bibnamefont {Medeiros}},\ and\ \bibinfo {author}
  {\bibfnamefont {K.~K.}\ \bibnamefont {Berggren}},\ }\bibfield  {title}
  {\bibinfo {title} {{}a scalable superconducting nanowire memory cell and
  preliminary array test},\ }\href {https://doi.org/10.1088/1361-6668/abd14e}
  {\bibfield  {journal} {\bibinfo  {journal} {Supercond. Sci. Technol.}\
  }\textbf {\bibinfo {volume} {34}},\ \bibinfo {pages} {035003} (\bibinfo
  {year} {2021})}\BibitemShut {NoStop}%
\bibitem [{\citenamefont {Hilgenkamp}(2021)}]{Hilgenkamp2020}%
  \BibitemOpen
  \bibfield  {author} {\bibinfo {author} {\bibfnamefont {H.}~\bibnamefont
  {Hilgenkamp}},\ }\bibfield  {title} {\bibinfo {title} {{Josephson
  Memories}},\ }\href {https://doi.org/10.1007/s10948-020-05680-2} {\bibfield
  {journal} {\bibinfo  {journal} {J. Supercond. Nov. Magn.}\ }\textbf {\bibinfo
  {volume} {34}},\ \bibinfo {pages} {1621} (\bibinfo {year}
  {2021})}\BibitemShut {NoStop}%
\bibitem [{\citenamefont {Larkin}\ \emph {et~al.}(2012)\citenamefont {Larkin},
  \citenamefont {Bol'ginov}, \citenamefont {Stolyarov}, \citenamefont
  {Ryazanov}, \citenamefont {Vernik}, \citenamefont {Tolpygo},\ and\
  \citenamefont {Mukhanov}}]{Larkin2012}%
  \BibitemOpen
  \bibfield  {author} {\bibinfo {author} {\bibfnamefont {T.~I.}\ \bibnamefont
  {Larkin}}, \bibinfo {author} {\bibfnamefont {V.~V.}\ \bibnamefont
  {Bol'ginov}}, \bibinfo {author} {\bibfnamefont {V.~S.}\ \bibnamefont
  {Stolyarov}}, \bibinfo {author} {\bibfnamefont {V.~V.}\ \bibnamefont
  {Ryazanov}}, \bibinfo {author} {\bibfnamefont {I.~V.}\ \bibnamefont
  {Vernik}}, \bibinfo {author} {\bibfnamefont {S.~K.}\ \bibnamefont
  {Tolpygo}},\ and\ \bibinfo {author} {\bibfnamefont {O.~A.}\ \bibnamefont
  {Mukhanov}},\ }\bibfield  {title} {\bibinfo {title} {{Ferromagnetic Josephson
  switching device with high characteristic voltage}},\ }\href
  {http://dx.doi.org/10.1063/1.4723576} {\bibfield  {journal} {\bibinfo
  {journal} {Appl. Phys. Lett.}\ }\textbf {\bibinfo {volume} {100}},\ \bibinfo
  {pages} {222601} (\bibinfo {year} {2012})}\BibitemShut {NoStop}%
\bibitem [{\citenamefont {Vernik}\ \emph {et~al.}(2013)\citenamefont {Vernik},
  \citenamefont {Bol'Ginov}, \citenamefont {Bakurskiy}, \citenamefont
  {Golubov}, \citenamefont {Kupriyanov}, \citenamefont {Ryazanov},\ and\
  \citenamefont {Mukhanov}}]{Vernik2013}%
  \BibitemOpen
  \bibfield  {author} {\bibinfo {author} {\bibfnamefont {I.~V.}\ \bibnamefont
  {Vernik}}, \bibinfo {author} {\bibfnamefont {V.~V.}\ \bibnamefont
  {Bol'Ginov}}, \bibinfo {author} {\bibfnamefont {S.~V.}\ \bibnamefont
  {Bakurskiy}}, \bibinfo {author} {\bibfnamefont {A.~A.}\ \bibnamefont
  {Golubov}}, \bibinfo {author} {\bibfnamefont {M.~Y.}\ \bibnamefont
  {Kupriyanov}}, \bibinfo {author} {\bibfnamefont {V.~V.}\ \bibnamefont
  {Ryazanov}},\ and\ \bibinfo {author} {\bibfnamefont {O.~A.}\ \bibnamefont
  {Mukhanov}},\ }\bibfield  {title} {\bibinfo {title} {{Magnetic Josephson
  junctions with superconducting interlayer for cryogenic memory}},\ }\href
  {https://doi.org/10.1109/TASC.2012.2233270} {\bibfield  {journal} {\bibinfo
  {journal} {IEEE Trans. Appl. Supercond.}\ }\textbf {\bibinfo {volume} {23}},\
  \bibinfo {pages} {1701208} (\bibinfo {year} {2013})}\BibitemShut {NoStop}%
\bibitem [{\citenamefont {Caruso}\ \emph
  {et~al.}(2018{\natexlab{a}})\citenamefont {Caruso}, \citenamefont
  {Massarotti}, \citenamefont {Bolginov}, \citenamefont {{Ben Hamida}},
  \citenamefont {Karelina}, \citenamefont {Miano}, \citenamefont {Vernik},
  \citenamefont {Tafuri}, \citenamefont {Ryazanov}, \citenamefont {Mukhanov},\
  and\ \citenamefont {Pepe}}]{Caruso2018}%
  \BibitemOpen
  \bibfield  {author} {\bibinfo {author} {\bibfnamefont {R.}~\bibnamefont
  {Caruso}}, \bibinfo {author} {\bibfnamefont {D.}~\bibnamefont {Massarotti}},
  \bibinfo {author} {\bibfnamefont {V.~V.}\ \bibnamefont {Bolginov}}, \bibinfo
  {author} {\bibfnamefont {A.}~\bibnamefont {{Ben Hamida}}}, \bibinfo {author}
  {\bibfnamefont {L.~N.}\ \bibnamefont {Karelina}}, \bibinfo {author}
  {\bibfnamefont {A.}~\bibnamefont {Miano}}, \bibinfo {author} {\bibfnamefont
  {I.~V.}\ \bibnamefont {Vernik}}, \bibinfo {author} {\bibfnamefont
  {F.}~\bibnamefont {Tafuri}}, \bibinfo {author} {\bibfnamefont {V.~V.}\
  \bibnamefont {Ryazanov}}, \bibinfo {author} {\bibfnamefont {O.~A.}\
  \bibnamefont {Mukhanov}},\ and\ \bibinfo {author} {\bibfnamefont {G.~P.}\
  \bibnamefont {Pepe}},\ }\bibfield  {title} {\bibinfo {title} {{RF assisted
  switching in magnetic Josephson junctions}},\ }\href
  {https://doi.org/10.1063/1.5018854} {\bibfield  {journal} {\bibinfo
  {journal} {J. Appl. Phys.}\ }\textbf {\bibinfo {volume} {123}},\ \bibinfo
  {pages} {133901} (\bibinfo {year} {2018}{\natexlab{a}})}\BibitemShut
  {NoStop}%
\bibitem [{\citenamefont {Caruso}\ \emph
  {et~al.}(2018{\natexlab{b}})\citenamefont {Caruso}, \citenamefont
  {Massarotti}, \citenamefont {Miano}, \citenamefont {Bolginov}, \citenamefont
  {Hamida}, \citenamefont {Karelina}, \citenamefont {Campagnano}, \citenamefont
  {Vernik}, \citenamefont {Tafuri}, \citenamefont {Ryazanov}, \citenamefont
  {Mukhanov},\ and\ \citenamefont {Pepe}}]{Caruso2018a}%
  \BibitemOpen
  \bibfield  {author} {\bibinfo {author} {\bibfnamefont {R.}~\bibnamefont
  {Caruso}}, \bibinfo {author} {\bibfnamefont {D.}~\bibnamefont {Massarotti}},
  \bibinfo {author} {\bibfnamefont {A.}~\bibnamefont {Miano}}, \bibinfo
  {author} {\bibfnamefont {V.~V.}\ \bibnamefont {Bolginov}}, \bibinfo {author}
  {\bibfnamefont {A.~B.}\ \bibnamefont {Hamida}}, \bibinfo {author}
  {\bibfnamefont {L.~N.}\ \bibnamefont {Karelina}}, \bibinfo {author}
  {\bibfnamefont {G.}~\bibnamefont {Campagnano}}, \bibinfo {author}
  {\bibfnamefont {I.~V.}\ \bibnamefont {Vernik}}, \bibinfo {author}
  {\bibfnamefont {F.}~\bibnamefont {Tafuri}}, \bibinfo {author} {\bibfnamefont
  {V.~V.}\ \bibnamefont {Ryazanov}}, \bibinfo {author} {\bibfnamefont {O.~A.}\
  \bibnamefont {Mukhanov}},\ and\ \bibinfo {author} {\bibfnamefont {G.~P.}\
  \bibnamefont {Pepe}},\ }\bibfield  {title} {\bibinfo {title} {{Properties of
  ferromagnetic josephson junctions for memory applications}},\ }\href
  {https://doi.org/10.1109/TASC.2018.2836979} {\bibfield  {journal} {\bibinfo
  {journal} {IEEE Trans. Appl. Supercond.}\ }\textbf {\bibinfo {volume} {28}},\
  \bibinfo {pages} {1800606} (\bibinfo {year}
  {2018}{\natexlab{b}})}\BibitemShut {NoStop}%
\bibitem [{\citenamefont {Baek}\ \emph {et~al.}(2014)\citenamefont {Baek},
  \citenamefont {Rippard}, \citenamefont {Benz}, \citenamefont {Russek},\ and\
  \citenamefont {Dresselhaus}}]{Baek2014}%
  \BibitemOpen
  \bibfield  {author} {\bibinfo {author} {\bibfnamefont {B.}~\bibnamefont
  {Baek}}, \bibinfo {author} {\bibfnamefont {W.~H.}\ \bibnamefont {Rippard}},
  \bibinfo {author} {\bibfnamefont {S.~P.}\ \bibnamefont {Benz}}, \bibinfo
  {author} {\bibfnamefont {S.~E.}\ \bibnamefont {Russek}},\ and\ \bibinfo
  {author} {\bibfnamefont {P.~D.}\ \bibnamefont {Dresselhaus}},\ }\bibfield
  {title} {\bibinfo {title} {{Hybrid superconducting-magnetic memory device
  using competing order parameters}},\ }\href
  {https://doi.org/10.1038/ncomms4888} {\bibfield  {journal} {\bibinfo
  {journal} {Nat. Commun.}\ }\textbf {\bibinfo {volume} {5}},\ \bibinfo {pages}
  {3888} (\bibinfo {year} {2014})}\BibitemShut {NoStop}%
\bibitem [{\citenamefont {Golod}\ \emph {et~al.}(2015)\citenamefont {Golod},
  \citenamefont {Iovan},\ and\ \citenamefont {Krasnov}}]{Golod2015}%
  \BibitemOpen
  \bibfield  {author} {\bibinfo {author} {\bibfnamefont {T.}~\bibnamefont
  {Golod}}, \bibinfo {author} {\bibfnamefont {A.}~\bibnamefont {Iovan}},\ and\
  \bibinfo {author} {\bibfnamefont {V.~M.}\ \bibnamefont {Krasnov}},\
  }\bibfield  {title} {\bibinfo {title} {{Single Abrikosov vortices as
  quantized information bits}},\ }\href {https://doi.org/10.1038/ncomms9628}
  {\bibfield  {journal} {\bibinfo  {journal} {Nat. Commun.}\ }\textbf {\bibinfo
  {volume} {6}},\ \bibinfo {pages} {8628} (\bibinfo {year} {2015})}\BibitemShut
  {NoStop}%
\bibitem [{\citenamefont {Dayton}\ \emph {et~al.}(2018)\citenamefont {Dayton},
  \citenamefont {Sage}, \citenamefont {Gingrich}, \citenamefont {Loving},
  \citenamefont {Ambrose}, \citenamefont {Siwak}, \citenamefont {Keebaugh},
  \citenamefont {Kirby}, \citenamefont {Miller}, \citenamefont {Herr},
  \citenamefont {Herr},\ and\ \citenamefont {Naaman}}]{Dayton2018}%
  \BibitemOpen
  \bibfield  {author} {\bibinfo {author} {\bibfnamefont {I.~M.}\ \bibnamefont
  {Dayton}}, \bibinfo {author} {\bibfnamefont {T.}~\bibnamefont {Sage}},
  \bibinfo {author} {\bibfnamefont {E.~C.}\ \bibnamefont {Gingrich}}, \bibinfo
  {author} {\bibfnamefont {M.~G.}\ \bibnamefont {Loving}}, \bibinfo {author}
  {\bibfnamefont {T.~F.}\ \bibnamefont {Ambrose}}, \bibinfo {author}
  {\bibfnamefont {N.~P.}\ \bibnamefont {Siwak}}, \bibinfo {author}
  {\bibfnamefont {S.}~\bibnamefont {Keebaugh}}, \bibinfo {author}
  {\bibfnamefont {C.}~\bibnamefont {Kirby}}, \bibinfo {author} {\bibfnamefont
  {D.~L.}\ \bibnamefont {Miller}}, \bibinfo {author} {\bibfnamefont {A.~Y.}\
  \bibnamefont {Herr}}, \bibinfo {author} {\bibfnamefont {Q.~P.}\ \bibnamefont
  {Herr}},\ and\ \bibinfo {author} {\bibfnamefont {O.}~\bibnamefont {Naaman}},\
  }\bibfield  {title} {\bibinfo {title} {{Experimental demonstration of a
  Josephson magnetic memory cell with a programmable $\pi$-junction}},\ }\href
  {https://doi.org/10.1109/LMAG.2018.2801820} {\bibfield  {journal} {\bibinfo
  {journal} {IEEE Magn. Lett.}\ }\textbf {\bibinfo {volume} {9}} (\bibinfo
  {year} {2018})}\BibitemShut {NoStop}%
\bibitem [{\citenamefont {Nevirkovets}\ and\ \citenamefont
  {Mukhanov}(2018)}]{Nevirkovets2018}%
  \BibitemOpen
  \bibfield  {author} {\bibinfo {author} {\bibfnamefont {I.~P.}\ \bibnamefont
  {Nevirkovets}}\ and\ \bibinfo {author} {\bibfnamefont {O.~A.}\ \bibnamefont
  {Mukhanov}},\ }\bibfield  {title} {\bibinfo {title} {Memory cell for
  high-density arrays based on a multiterminal superconducting-ferromagnetic
  device},\ }\href {https://doi.org/10.1103/PhysRevApplied.10.034013}
  {\bibfield  {journal} {\bibinfo  {journal} {Phys. Rev. Applied}\ }\textbf
  {\bibinfo {volume} {10}},\ \bibinfo {pages} {034013} (\bibinfo {year}
  {2018})}\BibitemShut {NoStop}%
\bibitem [{\citenamefont {Niedzielski}\ \emph {et~al.}(2018)\citenamefont
  {Niedzielski}, \citenamefont {Bertus}, \citenamefont {Glick}, \citenamefont
  {Loloee}, \citenamefont {Pratt},\ and\ \citenamefont
  {Birge}}]{Niedzielski2018}%
  \BibitemOpen
  \bibfield  {author} {\bibinfo {author} {\bibfnamefont {B.~M.}\ \bibnamefont
  {Niedzielski}}, \bibinfo {author} {\bibfnamefont {T.~J.}\ \bibnamefont
  {Bertus}}, \bibinfo {author} {\bibfnamefont {J.~A.}\ \bibnamefont {Glick}},
  \bibinfo {author} {\bibfnamefont {R.}~\bibnamefont {Loloee}}, \bibinfo
  {author} {\bibfnamefont {W.~P.}\ \bibnamefont {Pratt}},\ and\ \bibinfo
  {author} {\bibfnamefont {N.~O.}\ \bibnamefont {Birge}},\ }\bibfield  {title}
  {\bibinfo {title} {Spin-valve josephson junctions for cryogenic memory},\
  }\href {https://doi.org/10.1103/PhysRevB.97.024517} {\bibfield  {journal}
  {\bibinfo  {journal} {Phys. Rev. B}\ }\textbf {\bibinfo {volume} {97}},\
  \bibinfo {pages} {024517} (\bibinfo {year} {2018})}\BibitemShut {NoStop}%
\bibitem [{\citenamefont {Madden}\ \emph {et~al.}(2018)\citenamefont {Madden},
  \citenamefont {Willard}, \citenamefont {Loloee},\ and\ \citenamefont
  {Birge}}]{Madden2018}%
  \BibitemOpen
  \bibfield  {author} {\bibinfo {author} {\bibfnamefont {A.~E.}\ \bibnamefont
  {Madden}}, \bibinfo {author} {\bibfnamefont {J.~C.}\ \bibnamefont {Willard}},
  \bibinfo {author} {\bibfnamefont {R.}~\bibnamefont {Loloee}},\ and\ \bibinfo
  {author} {\bibfnamefont {N.~O.}\ \bibnamefont {Birge}},\ }\bibfield  {title}
  {\bibinfo {title} {Phase controllable josephson junctions for cryogenic
  memory},\ }\href {https://doi.org/10.1088/1361-6668/aae8cf} {\bibfield
  {journal} {\bibinfo  {journal} {Supercond. Sci. Technol.}\ }\textbf {\bibinfo
  {volume} {32}},\ \bibinfo {pages} {015001} (\bibinfo {year}
  {2018})}\BibitemShut {NoStop}%
\bibitem [{\citenamefont {Satchell}\ \emph {et~al.}(2020)\citenamefont
  {Satchell}, \citenamefont {Shepley}, \citenamefont {Algarni}, \citenamefont
  {Vaughan}, \citenamefont {Darwin}, \citenamefont {Ali}, \citenamefont
  {Rosamond}, \citenamefont {Chen}, \citenamefont {Linfield}, \citenamefont
  {Hickey},\ and\ \citenamefont {Burnell}}]{Satchell2020}%
  \BibitemOpen
  \bibfield  {author} {\bibinfo {author} {\bibfnamefont {N.}~\bibnamefont
  {Satchell}}, \bibinfo {author} {\bibfnamefont {P.~M.}\ \bibnamefont
  {Shepley}}, \bibinfo {author} {\bibfnamefont {M.}~\bibnamefont {Algarni}},
  \bibinfo {author} {\bibfnamefont {M.}~\bibnamefont {Vaughan}}, \bibinfo
  {author} {\bibfnamefont {E.}~\bibnamefont {Darwin}}, \bibinfo {author}
  {\bibfnamefont {M.}~\bibnamefont {Ali}}, \bibinfo {author} {\bibfnamefont
  {M.~C.}\ \bibnamefont {Rosamond}}, \bibinfo {author} {\bibfnamefont
  {L.}~\bibnamefont {Chen}}, \bibinfo {author} {\bibfnamefont {E.~H.}\
  \bibnamefont {Linfield}}, \bibinfo {author} {\bibfnamefont {B.~J.}\
  \bibnamefont {Hickey}},\ and\ \bibinfo {author} {\bibfnamefont
  {G.}~\bibnamefont {Burnell}},\ }\bibfield  {title} {\bibinfo {title}
  {Spin-valve josephson junctions with perpendicular magnetic anisotropy for
  cryogenic memory},\ }\href {https://doi.org/10.1063/1.5140095} {\bibfield
  {journal} {\bibinfo  {journal} {Appl. Phys. Lett.}\ }\textbf {\bibinfo
  {volume} {116}},\ \bibinfo {pages} {022601} (\bibinfo {year}
  {2020})}\BibitemShut {NoStop}%
\bibitem [{\citenamefont {Martinez}\ \emph {et~al.}(2016)\citenamefont
  {Martinez}, \citenamefont {Pratt},\ and\ \citenamefont
  {Birge}}]{Martinez2016}%
  \BibitemOpen
  \bibfield  {author} {\bibinfo {author} {\bibfnamefont {W.~M.}\ \bibnamefont
  {Martinez}}, \bibinfo {author} {\bibfnamefont {W.~P.}\ \bibnamefont
  {Pratt}},\ and\ \bibinfo {author} {\bibfnamefont {N.~O.}\ \bibnamefont
  {Birge}},\ }\bibfield  {title} {\bibinfo {title} {{Amplitude Control of the
  Spin-Triplet Supercurrent in $S/F/S$ Josephson Junctions}},\ }\href
  {https://doi.org/10.1103/PhysRevLett.116.077001} {\bibfield  {journal}
  {\bibinfo  {journal} {Phys. Rev. Lett.}\ }\textbf {\bibinfo {volume} {116}},\
  \bibinfo {pages} {077001} (\bibinfo {year} {2016})}\BibitemShut {NoStop}%
\bibitem [{\citenamefont {Glick}\ \emph {et~al.}(2018)\citenamefont {Glick},
  \citenamefont {Aguilar}, \citenamefont {Gougam}, \citenamefont {Niedzielski},
  \citenamefont {Gingrich}, \citenamefont {Loloee}, \citenamefont {Pratt},\
  and\ \citenamefont {Birge}}]{Glick2018}%
  \BibitemOpen
  \bibfield  {author} {\bibinfo {author} {\bibfnamefont {J.~A.}\ \bibnamefont
  {Glick}}, \bibinfo {author} {\bibfnamefont {V.}~\bibnamefont {Aguilar}},
  \bibinfo {author} {\bibfnamefont {A.~B.}\ \bibnamefont {Gougam}}, \bibinfo
  {author} {\bibfnamefont {B.~M.}\ \bibnamefont {Niedzielski}}, \bibinfo
  {author} {\bibfnamefont {E.~C.}\ \bibnamefont {Gingrich}}, \bibinfo {author}
  {\bibfnamefont {R.}~\bibnamefont {Loloee}}, \bibinfo {author} {\bibfnamefont
  {W.~P.}\ \bibnamefont {Pratt}},\ and\ \bibinfo {author} {\bibfnamefont
  {N.~O.}\ \bibnamefont {Birge}},\ }\bibfield  {title} {\bibinfo {title} {Phase
  control in a spin-triplet squid},\ }\href
  {https://doi.org/10.1126/sciadv.aat9457} {\bibfield  {journal} {\bibinfo
  {journal} {Sci. Adv.}\ }\textbf {\bibinfo {volume} {4}},\ \bibinfo {pages}
  {eaat9457} (\bibinfo {year} {2018})}\BibitemShut {NoStop}%
\bibitem [{\citenamefont {de~Andr\'es~Prada}\ \emph {et~al.}(2019)\citenamefont
  {de~Andr\'es~Prada}, \citenamefont {Golod}, \citenamefont {Kapran},
  \citenamefont {Borodianskyi}, \citenamefont {Bernhard},\ and\ \citenamefont
  {Krasnov}}]{DeAndresPrada2019}%
  \BibitemOpen
  \bibfield  {author} {\bibinfo {author} {\bibfnamefont {R.}~\bibnamefont
  {de~Andr\'es~Prada}}, \bibinfo {author} {\bibfnamefont {T.}~\bibnamefont
  {Golod}}, \bibinfo {author} {\bibfnamefont {O.~M.}\ \bibnamefont {Kapran}},
  \bibinfo {author} {\bibfnamefont {E.~A.}\ \bibnamefont {Borodianskyi}},
  \bibinfo {author} {\bibfnamefont {C.}~\bibnamefont {Bernhard}},\ and\
  \bibinfo {author} {\bibfnamefont {V.~M.}\ \bibnamefont {Krasnov}},\
  }\bibfield  {title} {\bibinfo {title} {{Memory-functionality
  superconductor/ferromagnet/superconductor junctions based on the
  high-${T}_{\text{c}}$ cuprate superconductors YBa$_{2}$Cu$_{3}$O$_{7-x}$ and
  the colossal magnetoresistive manganite ferromagnets
  La$_{2/3}$X$_{1/3}$MnO$_{3+\delta}$(X=Ca,Sr)}},\ }\href
  {https://doi.org/10.1103/PhysRevB.99.214510} {\bibfield  {journal} {\bibinfo
  {journal} {Phys. Rev. B}\ }\textbf {\bibinfo {volume} {99}},\ \bibinfo
  {pages} {214510} (\bibinfo {year} {2019})}\BibitemShut {NoStop}%
\bibitem [{\citenamefont {Weigand}\ \emph {et~al.}(2009)\citenamefont
  {Weigand}, \citenamefont {{Van Waeyenberge}}, \citenamefont {Vansteenkiste},
  \citenamefont {Curcic}, \citenamefont {Sackmann}, \citenamefont {Stoll},
  \citenamefont {Tyliszczak}, \citenamefont {Kaznatcheev}, \citenamefont
  {Bertwistle}, \citenamefont {Woltersdorf}, \citenamefont {Back},\ and\
  \citenamefont {Sch{\"{u}}tz}}]{Weigand2009}%
  \BibitemOpen
  \bibfield  {author} {\bibinfo {author} {\bibfnamefont {M.}~\bibnamefont
  {Weigand}}, \bibinfo {author} {\bibfnamefont {B.}~\bibnamefont {{Van
  Waeyenberge}}}, \bibinfo {author} {\bibfnamefont {A.}~\bibnamefont
  {Vansteenkiste}}, \bibinfo {author} {\bibfnamefont {M.}~\bibnamefont
  {Curcic}}, \bibinfo {author} {\bibfnamefont {V.}~\bibnamefont {Sackmann}},
  \bibinfo {author} {\bibfnamefont {H.}~\bibnamefont {Stoll}}, \bibinfo
  {author} {\bibfnamefont {T.}~\bibnamefont {Tyliszczak}}, \bibinfo {author}
  {\bibfnamefont {K.}~\bibnamefont {Kaznatcheev}}, \bibinfo {author}
  {\bibfnamefont {D.}~\bibnamefont {Bertwistle}}, \bibinfo {author}
  {\bibfnamefont {G.}~\bibnamefont {Woltersdorf}}, \bibinfo {author}
  {\bibfnamefont {C.~H.}\ \bibnamefont {Back}},\ and\ \bibinfo {author}
  {\bibfnamefont {G.}~\bibnamefont {Sch{\"{u}}tz}},\ }\bibfield  {title}
  {\bibinfo {title} {{Vortex core switching by coherent excitation with single
  in-plane magnetic field pulses}},\ }\href
  {https://doi.org/10.1103/PhysRevLett.102.077201} {\bibfield  {journal}
  {\bibinfo  {journal} {Phys. Rev. Lett.}\ }\textbf {\bibinfo {volume} {102}},\
  \bibinfo {pages} {077201} (\bibinfo {year} {2009})}\BibitemShut {NoStop}%
\bibitem [{\citenamefont {Uhl{\'{i}}{\v{r}}}\ \emph {et~al.}(2013)\citenamefont
  {Uhl{\'{i}}{\v{r}}}, \citenamefont {Urb{\'{a}}nek}, \citenamefont {Hlad{\'{i}}k},
  \citenamefont {Spousta}, \citenamefont {Im}, \citenamefont {Fischer},
  \citenamefont {Eibagi}, \citenamefont {Kan}, \citenamefont {Fullerton},\ and\
  \citenamefont {{\v{S}}ikola}}]{Uhlir2013}%
  \BibitemOpen
  \bibfield  {author} {\bibinfo {author} {\bibfnamefont {V.}~\bibnamefont
  {Uhl{\'{i}}{\v{r}}}}, \bibinfo {author} {\bibfnamefont {M.}~\bibnamefont
  {Urb{\'{a}}nek}}, \bibinfo {author} {\bibfnamefont {L.}~\bibnamefont
  {Hlad{\'{i}}k}}, \bibinfo {author} {\bibfnamefont {J.}~\bibnamefont
  {Spousta}}, \bibinfo {author} {\bibfnamefont {M.~Y.}\ \bibnamefont {Im}},
  \bibinfo {author} {\bibfnamefont {P.}~\bibnamefont {Fischer}}, \bibinfo
  {author} {\bibfnamefont {N.}~\bibnamefont {Eibagi}}, \bibinfo {author}
  {\bibfnamefont {J.~J.}\ \bibnamefont {Kan}}, \bibinfo {author} {\bibfnamefont
  {E.~E.}\ \bibnamefont {Fullerton}},\ and\ \bibinfo {author} {\bibfnamefont
  {T.}~\bibnamefont {{\v{S}}ikola}},\ }\bibfield  {title} {\bibinfo {title}
  {{Dynamic switching of the spin circulation in tapered magnetic nanodisks}},\
  }\href {https://doi.org/10.1038/nnano.2013.66} {\bibfield  {journal}
  {\bibinfo  {journal} {Nat. Nanotechnol.}\ }\textbf {\bibinfo {volume} {8}},\
  \bibinfo {pages} {341} (\bibinfo {year} {2013})}\BibitemShut {NoStop}%
\bibitem [{\citenamefont {Kammerer}\ \emph {et~al.}(2011)\citenamefont
  {Kammerer}, \citenamefont {Weigand}, \citenamefont {Curcic}, \citenamefont
  {Noske}, \citenamefont {Sproll}, \citenamefont {Vansteenkiste}, \citenamefont
  {{Van Waeyenberge}}, \citenamefont {Stoll}, \citenamefont {Woltersdorf},
  \citenamefont {Back},\ and\ \citenamefont {Schuetz}}]{Kammerer2011}%
  \BibitemOpen
  \bibfield  {author} {\bibinfo {author} {\bibfnamefont {M.}~\bibnamefont
  {Kammerer}}, \bibinfo {author} {\bibfnamefont {M.}~\bibnamefont {Weigand}},
  \bibinfo {author} {\bibfnamefont {M.}~\bibnamefont {Curcic}}, \bibinfo
  {author} {\bibfnamefont {M.}~\bibnamefont {Noske}}, \bibinfo {author}
  {\bibfnamefont {M.}~\bibnamefont {Sproll}}, \bibinfo {author} {\bibfnamefont
  {A.}~\bibnamefont {Vansteenkiste}}, \bibinfo {author} {\bibfnamefont
  {B.}~\bibnamefont {{Van Waeyenberge}}}, \bibinfo {author} {\bibfnamefont
  {H.}~\bibnamefont {Stoll}}, \bibinfo {author} {\bibfnamefont
  {G.}~\bibnamefont {Woltersdorf}}, \bibinfo {author} {\bibfnamefont {C.~H.}\
  \bibnamefont {Back}},\ and\ \bibinfo {author} {\bibfnamefont
  {G.}~\bibnamefont {Schuetz}},\ }\bibfield  {title} {\bibinfo {title}
  {{Magnetic vortex core reversal by excitation of spin waves}},\ }\href
  {https://doi.org/10.1038/ncomms1277} {\bibfield  {journal} {\bibinfo
  {journal} {Nat. Commun.}\ }\textbf {\bibinfo {volume} {2}},\ \bibinfo {pages}
  {279} (\bibinfo {year} {2011})}\BibitemShut {NoStop}%
\bibitem [{\citenamefont {Linder}\ and\ \citenamefont
  {Robinson}(2015)}]{Linder2015}%
  \BibitemOpen
  \bibfield  {author} {\bibinfo {author} {\bibfnamefont {J.}~\bibnamefont
  {Linder}}\ and\ \bibinfo {author} {\bibfnamefont {J.~W.~A.}\ \bibnamefont
  {Robinson}},\ }\bibfield  {title} {\bibinfo {title} {Superconducting
  spintronics},\ }\href {http://dx.doi.org/10.1038/nphys3242} {\bibfield
  {journal} {\bibinfo  {journal} {Nat. Phys.}\ }\textbf {\bibinfo {volume}
  {11}},\ \bibinfo {pages} {307} (\bibinfo {year} {2015})}\BibitemShut
  {NoStop}%
\bibitem [{\citenamefont {Eschrig}\ and\ \citenamefont
  {L{\"{o}}fwander}(2008)}]{Eschrig2008}%
  \BibitemOpen
  \bibfield  {author} {\bibinfo {author} {\bibfnamefont {M.}~\bibnamefont
  {Eschrig}}\ and\ \bibinfo {author} {\bibfnamefont {T.}~\bibnamefont
  {L{\"{o}}fwander}},\ }\bibfield  {title} {\bibinfo {title} {{Triplet
  supercurrents in clean and disordered half-metallic ferromagnets}},\ }\href
  {https://doi.org/10.1038/nphys831} {\bibfield  {journal} {\bibinfo  {journal}
  {Nat. Phys.}\ }\textbf {\bibinfo {volume} {4}},\ \bibinfo {pages} {138}
  (\bibinfo {year} {2008})}\BibitemShut {NoStop}%
\bibitem [{\citenamefont {Yang}\ \emph {et~al.}(2021)\citenamefont {Yang},
  \citenamefont {Ciccarelli},\ and\ \citenamefont {Robinson}}]{Robinson2021}%
  \BibitemOpen
  \bibfield  {author} {\bibinfo {author} {\bibfnamefont {G.}~\bibnamefont
  {Yang}}, \bibinfo {author} {\bibfnamefont {C.}~\bibnamefont {Ciccarelli}},\
  and\ \bibinfo {author} {\bibfnamefont {J.~W.~A.}\ \bibnamefont {Robinson}},\
  }\bibfield  {title} {\bibinfo {title} {{Boosting spintronics with
  superconductivity}},\ }\href {https://doi.org/10.1063/5.0048904} {\bibfield
  {journal} {\bibinfo  {journal} {APL Mater.}\ }\textbf {\bibinfo {volume}
  {9}},\ \bibinfo {pages} {050703} (\bibinfo {year} {2021})}\BibitemShut
  {NoStop}%
\bibitem [{\citenamefont {Lahabi}\ \emph {et~al.}(2017)\citenamefont {Lahabi},
  \citenamefont {Amundsen}, \citenamefont {Ouassou}, \citenamefont {Beukers},
  \citenamefont {Pleijster}, \citenamefont {Linder}, \citenamefont {Alkemade},\
  and\ \citenamefont {Aarts}}]{Lahabi2017a}%
  \BibitemOpen
  \bibfield  {author} {\bibinfo {author} {\bibfnamefont {K.}~\bibnamefont
  {Lahabi}}, \bibinfo {author} {\bibfnamefont {M.}~\bibnamefont {Amundsen}},
  \bibinfo {author} {\bibfnamefont {J.~A.}\ \bibnamefont {Ouassou}}, \bibinfo
  {author} {\bibfnamefont {E.}~\bibnamefont {Beukers}}, \bibinfo {author}
  {\bibfnamefont {M.}~\bibnamefont {Pleijster}}, \bibinfo {author}
  {\bibfnamefont {J.}~\bibnamefont {Linder}}, \bibinfo {author} {\bibfnamefont
  {P.}~\bibnamefont {Alkemade}},\ and\ \bibinfo {author} {\bibfnamefont
  {J.}~\bibnamefont {Aarts}},\ }\bibfield  {title} {\bibinfo {title}
  {{Controlling supercurrents and their spatial distribution in
  ferromagnets}},\ }\href {https://doi.org/10.1038/s41467-017-02236-2}
  {\bibfield  {journal} {\bibinfo  {journal} {Nat. Commun.}\ }\textbf {\bibinfo
  {volume} {8}},\ \bibinfo {pages} {2056} (\bibinfo {year} {2017})}\BibitemShut
  {NoStop}%
\bibitem [{\citenamefont {Fermin}\ \emph {et~al.}(2022)\citenamefont {Fermin},
  \citenamefont {van Dinter}, \citenamefont {Hubert}, \citenamefont {Woltjes},
  \citenamefont {Silaev}, \citenamefont {Aarts},\ and\ \citenamefont
  {Lahabi}}]{Co_disk_paper}%
  \BibitemOpen
  \bibfield  {author} {\bibinfo {author} {\bibfnamefont {R.}~\bibnamefont
  {Fermin}}, \bibinfo {author} {\bibfnamefont {D.}~\bibnamefont {van Dinter}},
  \bibinfo {author} {\bibfnamefont {M.}~\bibnamefont {Hubert}}, \bibinfo
  {author} {\bibfnamefont {B.}~\bibnamefont {Woltjes}}, \bibinfo {author}
  {\bibfnamefont {M.}~\bibnamefont {Silaev}}, \bibinfo {author} {\bibfnamefont
  {J.}~\bibnamefont {Aarts}},\ and\ \bibinfo {author} {\bibfnamefont
  {K.}~\bibnamefont {Lahabi}},\ }\bibfield  {title} {\bibinfo {title}
  {Superconducting triplet rim currents in a spin-textured ferromagnetic
  disk},\ }\href {https://doi.org/10.1021/acs.nanolett.1c04051} {\bibfield
  {journal} {\bibinfo  {journal} {Nano Letters}\ }\textbf {\bibinfo {volume}
  {22}},\ \bibinfo {pages} {2209} (\bibinfo {year} {2022})}\BibitemShut
  {NoStop}%
\bibitem [{Note1()}]{Note1}%
  \BibitemOpen
  \bibinfo {note} {Samples with higher $I_{\protect \text {c}}$ values do show
  a Fraunhofer interference pattern, indicating a uniform distribution of
  $I_{\protect \text {c}}$ throughout the weak link.}\BibitemShut {Stop}%
\bibitem [{\citenamefont {Singh}\ \emph {et~al.}(2016)\citenamefont {Singh},
  \citenamefont {Jansen}, \citenamefont {Lahabi},\ and\ \citenamefont
  {Aarts}}]{Singh2016}%
  \BibitemOpen
  \bibfield  {author} {\bibinfo {author} {\bibfnamefont {A.}~\bibnamefont
  {Singh}}, \bibinfo {author} {\bibfnamefont {C.}~\bibnamefont {Jansen}},
  \bibinfo {author} {\bibfnamefont {K.}~\bibnamefont {Lahabi}},\ and\ \bibinfo
  {author} {\bibfnamefont {J.}~\bibnamefont {Aarts}},\ }\bibfield  {title}
  {\bibinfo {title} {{High-quality ${\mathrm{CrO}}_{2}$ nanowires for
  dissipation-less spintronics}},\ }\href
  {https://link.aps.org/doi/10.1103/PhysRevX.6.041012} {\bibfield  {journal}
  {\bibinfo  {journal} {Phys. Rev. X}\ }\textbf {\bibinfo {volume} {6}},\
  \bibinfo {pages} {041012} (\bibinfo {year} {2016})}\BibitemShut {NoStop}%
\bibitem [{\citenamefont {Keizer}\ \emph {et~al.}(2006)\citenamefont {Keizer},
  \citenamefont {Goennenwein}, \citenamefont {Klapwijk}, \citenamefont {Miao},
  \citenamefont {Xiao},\ and\ \citenamefont {Gupta}}]{Keizer2006}%
  \BibitemOpen
  \bibfield  {author} {\bibinfo {author} {\bibfnamefont {R.~S.}\ \bibnamefont
  {Keizer}}, \bibinfo {author} {\bibfnamefont {S.~T.~B.}\ \bibnamefont
  {Goennenwein}}, \bibinfo {author} {\bibfnamefont {T.~M.}\ \bibnamefont
  {Klapwijk}}, \bibinfo {author} {\bibfnamefont {G.}~\bibnamefont {Miao}},
  \bibinfo {author} {\bibfnamefont {G.}~\bibnamefont {Xiao}},\ and\ \bibinfo
  {author} {\bibfnamefont {A.}~\bibnamefont {Gupta}},\ }\bibfield  {title}
  {\bibinfo {title} {{A spin triplet supercurrent through the half-metallic
  ferromagnet CrO$_2$}},\ }\href {http://dx.doi.org/10.1038/nature04499}
  {\bibfield  {journal} {\bibinfo  {journal} {Nature}\ }\textbf {\bibinfo
  {volume} {439}},\ \bibinfo {pages} {825} (\bibinfo {year}
  {2006})}\BibitemShut {NoStop}%
\bibitem [{\citenamefont {Anwar}\ \emph {et~al.}(2010)\citenamefont {Anwar},
  \citenamefont {Czeschka}, \citenamefont {Hesselberth}, \citenamefont
  {Porcu},\ and\ \citenamefont {Aarts}}]{Anwar2010}%
  \BibitemOpen
  \bibfield  {author} {\bibinfo {author} {\bibfnamefont {M.~S.}\ \bibnamefont
  {Anwar}}, \bibinfo {author} {\bibfnamefont {F.}~\bibnamefont {Czeschka}},
  \bibinfo {author} {\bibfnamefont {M.}~\bibnamefont {Hesselberth}}, \bibinfo
  {author} {\bibfnamefont {M.}~\bibnamefont {Porcu}},\ and\ \bibinfo {author}
  {\bibfnamefont {J.}~\bibnamefont {Aarts}},\ }\bibfield  {title} {\bibinfo
  {title} {{Long-range supercurrents through half-metallic ferromagnetic
  CrO$_2$}},\ }\href {https://link.aps.org/doi/10.1103/PhysRevB.82.100501}
  {\bibfield  {journal} {\bibinfo  {journal} {Phys. Rev. B}\ }\textbf {\bibinfo
  {volume} {82}},\ \bibinfo {pages} {100501(R)} (\bibinfo {year}
  {2010})}\BibitemShut {NoStop}%
\bibitem [{Note2()}]{Note2}%
  \BibitemOpen
  \bibinfo {note} {The length of the trench in the disk-shaped devices is
  typically between 1~$\mu $m and 1.6~$\mu $m, whereas the trench length of the
  ellipse devices presenter here is less than 600 nm.}\BibitemShut {Stop}%
\bibitem [{\citenamefont {Song}\ \emph {et~al.}(2022)\citenamefont {Song},
  \citenamefont {Chen}, \citenamefont {Hou}, \citenamefont {Qin}, \citenamefont
  {Gao},\ and\ \citenamefont {Liu}}]{Song2022}%
  \BibitemOpen
  \bibfield  {author} {\bibinfo {author} {\bibfnamefont {X.}~\bibnamefont
  {Song}}, \bibinfo {author} {\bibfnamefont {J.-P.}\ \bibnamefont {Chen}},
  \bibinfo {author} {\bibfnamefont {Z.-P.}\ \bibnamefont {Hou}}, \bibinfo
  {author} {\bibfnamefont {M.-H.}\ \bibnamefont {Qin}}, \bibinfo {author}
  {\bibfnamefont {X.-S.}\ \bibnamefont {Gao}},\ and\ \bibinfo {author}
  {\bibfnamefont {J.-M.}\ \bibnamefont {Liu}},\ }\bibfield  {title} {\bibinfo
  {title} {{Strain-mediated voltage-controlled magnetic double-vortex states in
  elliptical nanostructures}},\ }\href
  {https://doi.org/10.1016/j.jmmm.2021.168729} {\bibfield  {journal} {\bibinfo
  {journal} {J. Magn. Magn. Mater.}\ }\textbf {\bibinfo {volume} {547}},\
  \bibinfo {pages} {168729} (\bibinfo {year} {2022})}\BibitemShut {NoStop}%
\bibitem [{\citenamefont {Vansteenkiste}\ \emph {et~al.}(2014)\citenamefont
  {Vansteenkiste}, \citenamefont {Leliaert}, \citenamefont {Dvornik},
  \citenamefont {Helsen}, \citenamefont {Garcia-Sanchez},\ and\ \citenamefont
  {{Van Waeyenberge}}}]{Vansteenkiste2014}%
  \BibitemOpen
  \bibfield  {author} {\bibinfo {author} {\bibfnamefont {A.}~\bibnamefont
  {Vansteenkiste}}, \bibinfo {author} {\bibfnamefont {J.}~\bibnamefont
  {Leliaert}}, \bibinfo {author} {\bibfnamefont {M.}~\bibnamefont {Dvornik}},
  \bibinfo {author} {\bibfnamefont {M.}~\bibnamefont {Helsen}}, \bibinfo
  {author} {\bibfnamefont {F.}~\bibnamefont {Garcia-Sanchez}},\ and\ \bibinfo
  {author} {\bibfnamefont {B.}~\bibnamefont {{Van Waeyenberge}}},\ }\bibfield
  {title} {\bibinfo {title} {{The design and verification of MuMax3}},\ }\href
  {https://doi.org/10.1063/1.4899186} {\bibfield  {journal} {\bibinfo
  {journal} {AIP Adv.}\ }\textbf {\bibinfo {volume} {4}},\ \bibinfo {pages}
  {107133} (\bibinfo {year} {2014})}\BibitemShut {NoStop}%
\end{thebibliography}

%

\end{document}


\title{Supplementary Information of: Mesoscopic superconducting memory based on bistable magnetic textures}

\date{\today}

\author{R. Fermin}
\author{N. Scheinowitz}
\author{J. Aarts}
\affiliation{Huygens-Kamerlingh Onnes Laboratory, Leiden University, P.O. Box 9504, 2300 RA Leiden, The Netherlands.}
\author{K. Lahabi}
\affiliation{Huygens-Kamerlingh Onnes Laboratory, Leiden University, P.O. Box 9504, 2300 RA Leiden, The Netherlands.}\affiliation{Department of Quantum Nanoscience, Kavli Institute of Nanoscience, Delft University of Technology, Delft, 2628 CJ, The Netherlands}

\pacs{} \maketitle
 
\clearpage

\section{Additional data}

\begin{figure*}[h]
\centerline{$
\begin{array}{c}
\includegraphics[width=1\linewidth]{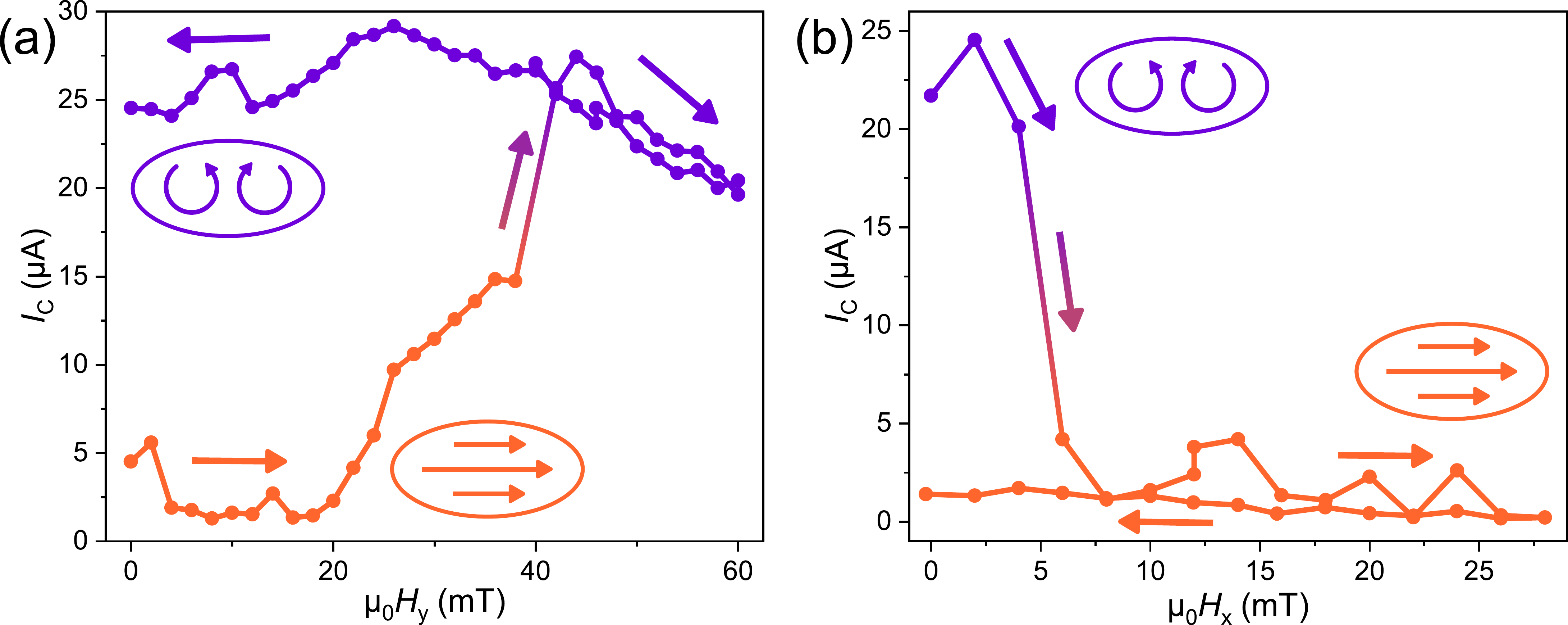}
\end{array}$}
\caption{The critical current $I_{\text{c}}$ as a function of H$_x$ (field along the long axis), and $H_y$ (field along the short axis). Similar data as presented in Figure 3 of the main text, obtained on device A at 1.7 K. (a) and (b) depict $I_{\text{c}}(H_y)$ and $I_{\text{c}}(H_x)$ respectively. In (a) the field is increased along the short axis of the ellipse. Like sample B, at 40~mT $I_\text{c}$ increases sharply, effectively transforming the M-state into the V-state. By increasing the field parallel to the long axis of the ellipse (shown in (b)), the critical current drops to the low $I_{\text{c}}$ state at 5 mT. This is lower than the 15 mT, needed to switch device B.}\label{Sfig1}
\end{figure*}

\begin{figure*}[tb]
\centerline{$
\begin{array}{c}
\includegraphics[width=0.6\linewidth]{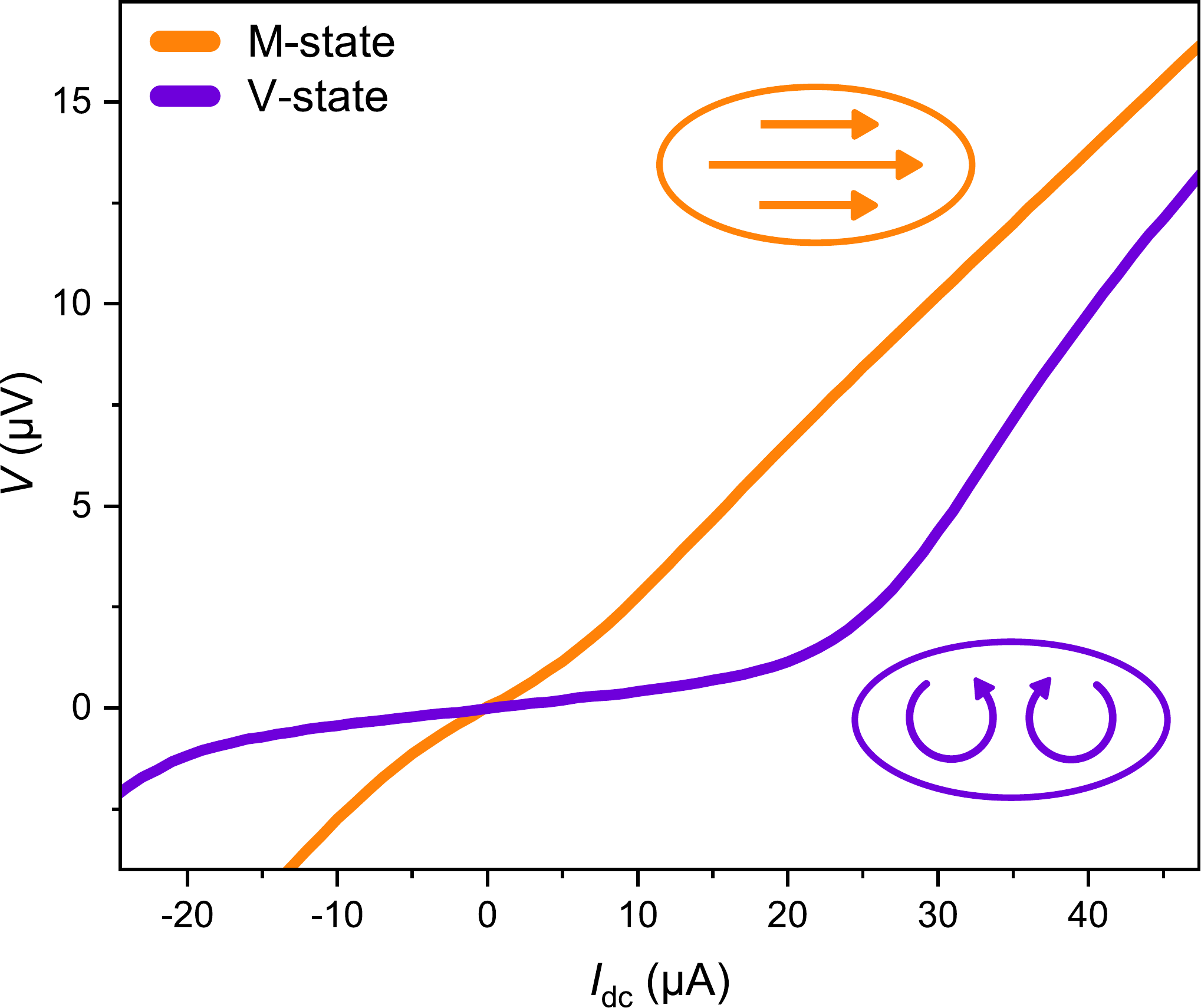}
\end{array}$}
\caption{The typical IV for both states in device B. By choosing a proper read-out current (e.g. 15 $\mu$A), a clear voltage difference between the two states can be measured.}\label{Sfig3}
\end{figure*}

\begin{figure*}[h]
\centerline{$
\begin{array}{c}
\includegraphics[width=1\linewidth]{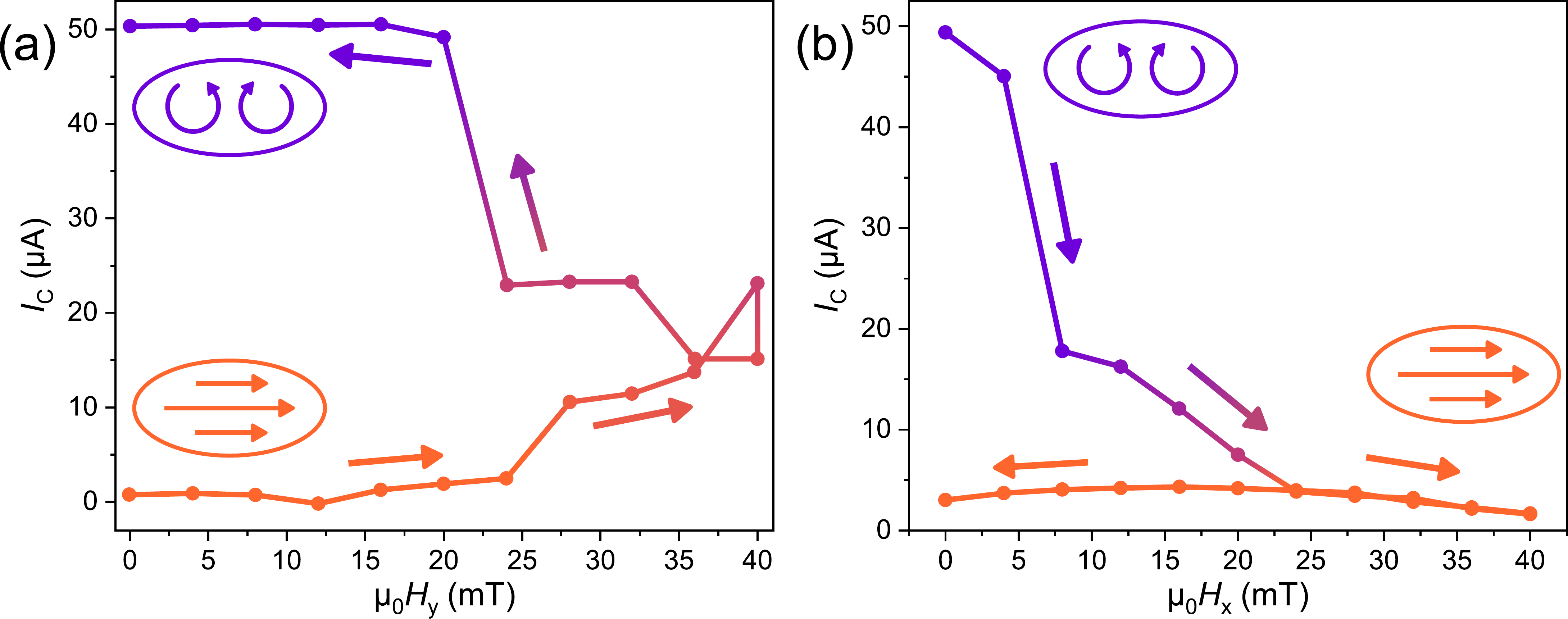}
\end{array}$}
\caption{The critical current $I_{\text{c}}$ as function of $H_y$ (field along the short axis, shown in (a)), and H$_x$ (field along the long axis, shown in (b)). Similar data as presented in Figure 3 of the main text, obtained on a third device, called device C. Data acquired at 2 K. In (a) the field is increased along the short axis of the ellipse; by increasing the field to 40 mT, the spin texture is buckled and upon reducing the field to zero, the V-state and entailing high $I_{\text{c}}$ state is retrieved. By increasing the field parallel to the long axis of the ellipse , the critical current drops to the low $I_{\text{c}}$ state corresponding to the M-state.}\label{Sfig5}
\end{figure*}

\begin{figure*}[tb]
\centerline{$
\begin{array}{c}
\includegraphics[width=1\linewidth]{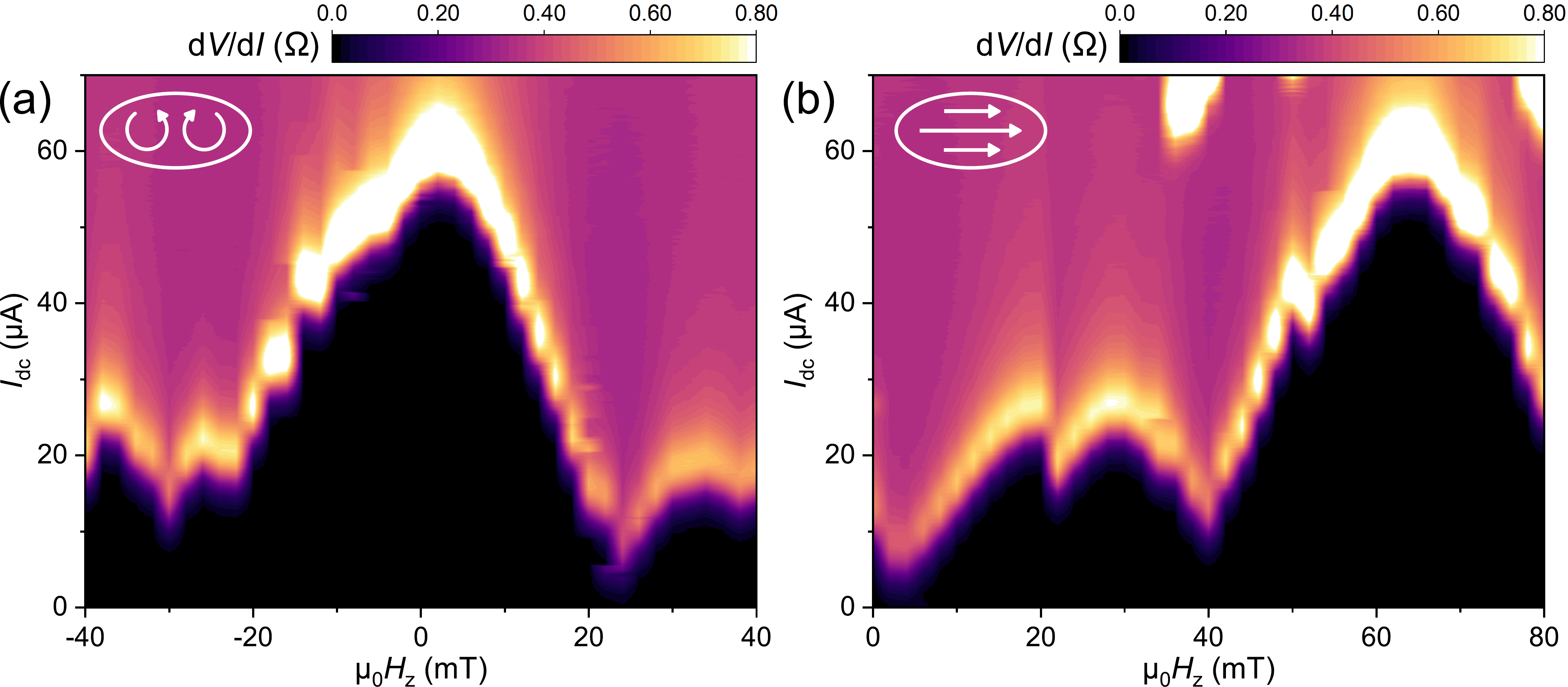}
\end{array}$}
\caption{(a) to (b) respectively show the superconducting interference (SQI) patterns in the V-state and the M-state in a third device, named device C. Although the shape of the SQI pattern does not change, the center lobe of the pattern is shifted in the V-state with respect to the pattern in the M-state. This is the same behavior as observed for device B (presented in the main text).}\label{Sfig6}
\end{figure*}

\begin{figure*}[tb]
\centerline{$
\begin{array}{c}
\includegraphics[width=1\linewidth]{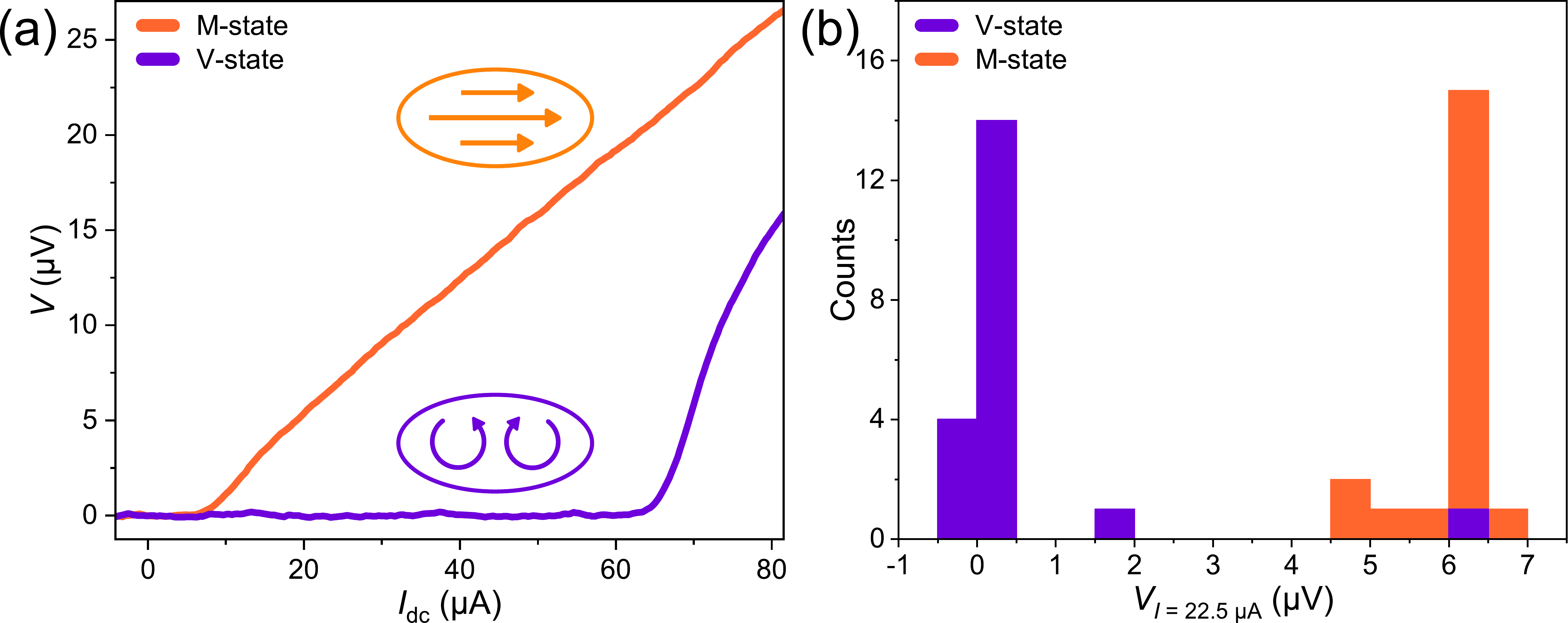}
\end{array}$}
\caption{Fidelity of writing cycles in a third device, named device C. In (a) we show the $IV$-characteristic for both states. We define a read-out current of 22.5 $\mu$A. We cycle the device between the states 20 times and record the voltage at the read-out current. (b) shows a histogram of these voltages.}\label{Sfig7}
\end{figure*}

In the main text we provide results on two different devices, named A and B. In this section we provide additional data obtained on these samples. Supporting Figure \ref{Sfig1} shows the in-plane field sweeps obtained on device A. Similarly as in device B, there is a transition from the M-state to the V- state at 40 mT, when increasing the field along the short axis of the ellipse. This is accompanied by an increase in critical current, which is retained upon reducing the field to zero. The converse operation is carried out by applying the field along the long axis of the ellipse. We find that 5 mT is sufficient to reduce the critical current to the value corresponding to the M-state. This is lower than the corresponding field for device B (15 mT). Figure \ref{Sfig3} shows the typical $IV$-characteristics of the V-state and M-state for device B. A clear voltage difference is measured at a read-out current of, for example, 15 $\mu$A. No histogram was obtained on this device. Finally, we present a full dataset obtained on a third device in Figures \ref{Sfig5},\ref{Sfig6} and, \ref{Sfig7}.

\clearpage

\section{Calculating the stray fields}

In Figure 5d of the main text we have provided calculations of the stray field that penetrates our SFS-junctions; here we provide further details on these calculations. In all our simulations we have included vacuum cells around the magnetic structure. These feature no spin, yet they can be penetrated by magnetic fields. In each steady state solution we thereby gain full insight into the stray fields around the magnetic layer. In order to calculate the stray fields penetrating the junction, we sum the z-component of these stray fields directly above the magnetic layer (i.e., penetrating the superconducting layer perpendicular to the charge transport direction). We define a window of area $A$ and find the field penetrating this window by dividing the flux ($\Phi$) by its area. 

\begin{figure*}[h]
\centerline{$
\begin{array}{c}
\includegraphics[width=0.7\linewidth]{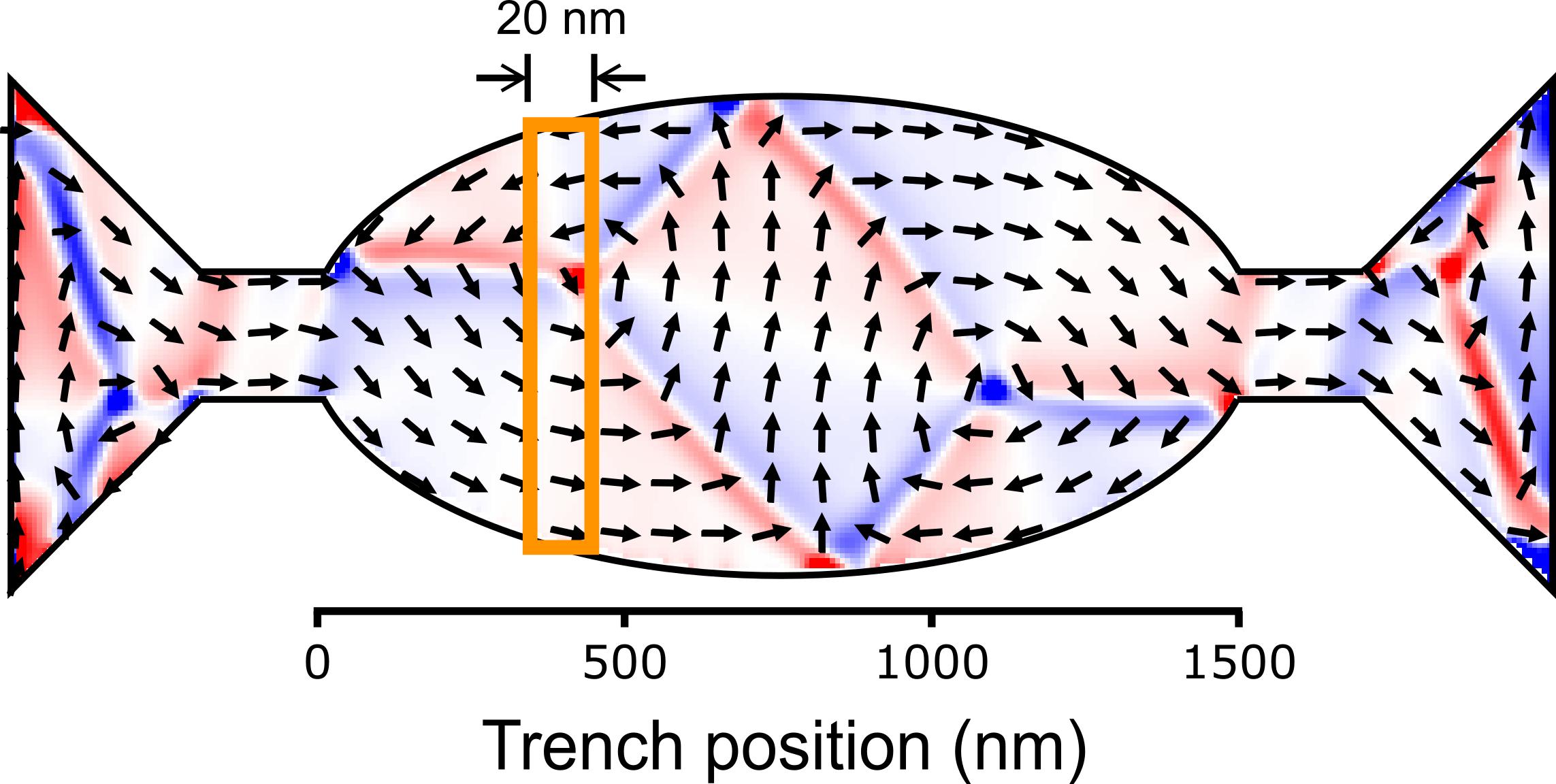}
\end{array}$}
\caption{Graphical representation of the integration window, used to calculate the local stray fields (in magenta). The integration window is chosen to reflect the size of the physical junction. The width equals that of the trench (20 nm) and the length is  equal to the local width of the ellipse. By positioning the integration window at different locations (i.e., trench positions), we can evaluate the stray fields at different locations along the ellipse. The axis labeled 'Trench position' coincides with the x-axis of Figure 5d of the main text.}\label{Sfigwindow}
\end{figure*}   

\noindent The flux can be found by summing the product of the field in each cell and the area of each cell. Since each cell has the same area by design, this leads to:

\begin{align}
\Delta B = \frac{\Phi}{A} = \frac{1}{A} \sum B_{z,\text{i}} \: A_{\text{cell,i}} = \frac{1}{N_{\text{cells}}}\sum B_{z,\text{i}}
\end{align}

\noindent Here we recognize that the total field penetrating the window equals the average field per cell, which can be evaluated from the simulations.

Ideally the area of the integration window is chosen such that it reflects the effective area of the junction. The planar junctions discussed here are in the thin film limit (i.e., the thickness of the superconducting layer is thinner than the London penetration depth $\lambda$, for all temperatures that are probed here). In this case the effective area becomes dependent on the geometry of the electrodes and not, as is the case for macroscopic junctions, dependent on $\lambda$. John Clem devised a general method for calculating the effective area for planar junctions in a constricted geometry.\cite{Clem2010} Since our devices feature asymmetric electrodes and a non-conventional geometry, evaluating the effective area would require a separate calculation for each junction position. Moreover, due to the tapered shape of the ellipse, the effective area is non-rectangular, further increasing the difficulty of establishing the stray fields. In order to solve these issues we specify a rectangular window with a fixed width (set by the junction width: 20 nm) and a variable length. We match the length of the window to that of the local width of the ellipse. By this choice, we minimize the summation of field outside the device geometry. By varying the location of the center of the integration window, we evaluate an expression of the stray fields that locally penetrate the superconductor. A graphical representation of the integration window is shown in Figure \ref{Sfigwindow}.